\documentclass[main,longauth]{aa}
\usepackage{graphicx}
\usepackage{amsmath}
\usepackage{txfonts}
\usepackage{cprotect}
\usepackage{orcidlink}

\def\Msun{M$_\odot$}
\def\Zsun{Z$_\odot$}

\def\Ha{H$\alpha$}
\def\Hb{H$\beta$}

\def\Heii{He\,{\sc ii}}

\def\Cii{[C\,{\sc ii}]}

\def\Nii{[N\,{\sc ii}]}

\def\Oi{[O\,{\sc i}]}
\def\Oii{[O\,{\sc ii}]}
\def\Oiii{[O\,{\sc iii}]}
\def\Sii{[S\,{\sc ii}]}
\def\kms{km\,s$^{-1}$}

\def\lsim{\mathrel{\rlap{\lower 3pt \hbox{$\sim$}} \raise 2.0pt \hbox{$<$}}}
\def\gsim{\mathrel{\rlap{\lower 3pt \hbox{$\sim$}} \raise 2.0pt \hbox{$>$}}}

\begin{document}

\authorrunning{Decarli et al.}
\titlerunning{Mapping a quasar+satellites merger at $z\sim6.2$}

\title{A quasar-galaxy merger at $z\sim 6.2$: rapid host growth via accretion of two massive satellite galaxies}

\author{
Roberto Decarli\inst{1}$^{\orcidlink{0000-0002-2662-8803}}$, 
Federica Loiacono\inst{1}$^{\orcidlink{0000-0002-8858-6784}}$, 
Emanuele Paolo Farina\inst{2}$^{\orcidlink{0000-0002-6822-2254}}$, 
Massimo Dotti\inst{3, 4, 5}$^{\orcidlink{0000-0002-1683-5198}}$, 
Alessandro Lupi\inst{6, 3, 4}$^{\orcidlink{0000-0001-6106-7821}}$, 
Romain A.~Meyer\inst{7}$^{\orcidlink{0000-0001-5492-4522}}$, 
Marco Mignoli\inst{1}$^{\orcidlink{0000-0002-9087-2835}}$, 
Antonio Pensabene\inst{3}$^{\orcidlink{0000-0001-9815-4953}}$, 
Michael A.~Strauss\inst{8}$^{\orcidlink{0000-0002-0106-7755}}$, 
Bram Venemans\inst{9}$^{\orcidlink{0000-0001-9024-8322}}$, 
Jinyi Yang\inst{10}$^{\orcidlink{0000-0001-5287-4242}}$
Fabian Walter\inst{11}$^{\orcidlink{0000-0003-4793-7880}}$, 
Julien Wolf\inst{11}$^{\orcidlink{0000-0003-0643-7935}}$,
Eduardo Ba\~{n}ados\inst{11}$^{\orcidlink{0000-0002-2931-7824}}$, 
Laura Blecha\inst{12}$^{\orcidlink{0000-0002-2183-1087}}$, 
Sarah Bosman\inst{13, 11}$^{\orcidlink{0000-0001-8582-7012}}$, 
Chris L.~Carilli\inst{14}$^{\orcidlink{0000-0001-6647-3861}}$, 
Andrea Comastri\inst{1}$^{\orcidlink{0000-0003-3451-9970}}$, 
Thomas Connor\inst{15, 16}$^{\orcidlink{0000-0002-7898-7664}}$, 
Tiago Costa\inst{17, 18}$^{\orcidlink{0000-0002-6748-2900}}$, 
Anna-Christina Eilers\inst{19}$^{\orcidlink{0000-0003-2895-6218}}$, 
Xiaohui Fan\inst{10}$^{\orcidlink{0000-0003-3310-0131}}$, 
Roberto Gilli\inst{1}$^{\orcidlink{0000-0001-8121-6177}}$, 
Hyunsung D.~Jun\inst{20}$^{\orcidlink{0000-0003-1470-5901}}$, 
Weizhe Liu\inst{10}$^{\orcidlink{0000-0003-3762-7344}}$, 
Madeline A.~Marshall\inst{21, 22}$^{\orcidlink{0000-0001-6434-7845}}$, 
Chiara Mazzucchelli\inst{23}$^{\orcidlink{0000-0002-5941-5214}}$, 
Marcel Neeleman\inst{14}$^{\orcidlink{0000-0002-9838-8191}}$, 
Masafusa Onoue\inst{24, 25, 26}$^{\orcidlink{0000-0003-2984-6803}}$
Roderik Overzier\inst{9}$^{\orcidlink{0000-0002-8214-7617}}$, 
Maria Anne Pudoka\inst{10}$^{\orcidlink{0000-0003-4924-5941}}$
Dominik A.~Riechers\inst{27}$^{\orcidlink{0000-0001-9585-1462}}$, 
Hans-Walter Rix\inst{11}$^{\orcidlink{0000-0003-4996-9069}}$
Jan-Torge Schindler\inst{28}$^{\orcidlink{0000-0002-4544-8242}}$, 
Benny Trakhtenbrot\inst{29}$^{\orcidlink{0000-0002-3683-7297}}$, 
Maxime Trebitsch\inst{30}$^{\orcidlink{0000-0002-6849-5375}}$, 
Marianne Vestergaard\inst{31, 10}$^{\orcidlink{0000-0001-9191-9837}}$, 
Marta Volonteri\inst{32}$^{\orcidlink{0000-0002-3216-1322}}$, 
Feige Wang\inst{10}$^{\orcidlink{0000-0002-7633-431X}}$, 
Huanian Zhang\inst{33}$^{\orcidlink{0000-0002-0123-9246}}$,
Siwei Zou\inst{34}$^{\orcidlink{0000-0002-3983-6484}}$
}
\institute{
INAF -- Osservatorio di Astrofisica e Scienza dello Spazio di Bologna, via Gobetti 93/3, I-40129, Bologna, Italy. \email{ roberto.decarli@inaf.it} \and 
Gemini Observatory, NSF’s NOIRLab, 670 N A’ohoku Place, Hilo, Hawai’i 96720, USA \and
Dipartimento di Fisica ``G. Occhialini'', Universit\`{a} degli Studi di Milano-Bicocca, Piazza della Scienza 3, I-20126 Milano, Italy \and 
INFN, Sezione di Milano-Bicocca, Piazza della Scienza 3, I-20126 Milano, Italy \and 
INAF - Osservatorio Astronomico di Brera, via Brera 20, I-20121 Milano, Italy \and 
Dipartimento di Scienza e Alta Tecnologia, Universit\`{a} degli Studi dell’Insubria, via Valleggio 11, I-22100, Como, Italy \and
Departement d'Astronomie, University of Geneva, Chemin Pegasi 51, 1290 Versoix, Switzerland \and 
Department of Astrophysical Sciences, Princeton University, Princeton, NJ 08544 USA \and 
Leiden Observatory, Leiden University, Niels Bohrweg 2, NL-2333 CA Leiden, Netherlands \and 
Steward Observatory, University of Arizona, 933 N Cherry Avenue, Tucson, AZ 85721, US \and 
Max Planck Institut f\"{u}r Astronomie, K\"{o}nigstuhl 17, D-69117, Heidelberg, Germany \and 
Department of Physics, University of Florida, Gainesville, FL 32611-8440, USA \and 
Institute for Theoretical Physics, University of Heidelberg, Philosophenweg 16, 69120 Heidelberg, Germany \and 
National Radio Astronomy Observatory, P.O. Box O, Socorro, NM 87801, USA \and 
Center for Astrophysics -- Harvard \& Smithsonian, 60 Garden St., Cambridge, MA 02138, USA \and 
Jet Propulsion Laboratory, California Institute of Technology, 4800 Oak Grove Drive, Pasadena, CA 91109, USA \and 
Max-Planck-Institut f\"{u}r Astrophysik, Karl-Schwarzschild-Straße 1, D-85748 Garching b. M\"{u}nchen, Germany \and 
Newcastle University, School of Mathematics, Statistics and Physics, Herschel Building, Newcastle upon Tyne NE1 7RU, United Kingdom \and
MIT Kavli Institute for Astrophysics and Space Research, 77 Massachusetts Ave., Cambridge, MA 02139, USA \and 
Department of Physics, Northwestern College, 101 7th St SW, Orange City, IA 51041, USA \and 
National Research Council of Canada, Herzberg Astronomy \& Astrophysics Research Centre, 5071 West Saanich Road, Victoria, BC V9E 2E7, Canada \and 
ARC Centre of Excellence for All Sky Astrophysics in 3 Dimensions, Australia \and 
Instituto de Estudios Astrof\'isicos, Facultad de Ingenier\'ia y Ciencias, Universidad Diego Portales, Avenida Ejercito Libertador 441, Santiago, Chile \and 
Kavli Institute for the Physics and Mathematics of the Universe (Kavli IPMU, WPI), The University of Tokyo, 5-1-5 Kashiwanoha, Kashiwa, Chiba 277-8583, Japan \and
Center for Data-Driven Discovery, Kavli IPMU (WPI), UTIAS, The University of Tokyo, Kashiwa, Chiba 277-8583, Japan\and
Kavli Institute for Astronomy and Astrophysics, Peking University, Beijing 100871, P.R.China \and
I. Physikalisches Institut, Universit\"{a}t zu K\"{o}ln, Z\"{u}lpicher Strasse 77, 50937 K\"{o}ln, Germany \and 
Hamburger Sternwarte, Universit\"{a}t Hamburg, Gojenbergsweg 112, D-21029 Hamburg, Germany \and 
School of Physics and Astronomy, Tel Aviv University, Tel Aviv 69978, Israel \and 
Kapteyn Astronomical Institute, University of Groningen, P.O. Box 800, 9700 AV Groningen, The Netherlands \and 
The Niels Bohr Institute, University of Copenhagen, Denmark \and 
Institut d’Astrophysique de Paris, Sorbonne Universit\'{e}, CNRS, UMR 7095, 98 bis bd Arago, 75014 Paris, France \and
Department of Astronomy, Huazhong University of Science and Technology, Wuhan, 430074, People's Republic of China \and
Department of Astronomy, Tsinghua University, Beijing 100084, People's Republic of China
}

\date{January 2024}

\abstract{
We present JWST/NIRSpec Integral Field Spectroscopy in the rest-frame optical bands of the system PJ308--21, a quasar at $z=6.2342$ caught as its host galaxy interacts with companion galaxies. We detect spatially extended emission of several emission lines (\Ha{}, \Hb{}, \Oiii{}, \Nii{}, \Sii{}, \Heii{}), which we use to study the properties of the ionized phase of the interstellar medium: the source and hardness of the photoionizing radiation field, metallicity, dust reddening, electron density and temperature, and star formation. We also marginally detect continuum starlight emission associated with the companion sources. We find that at least two independent satellite galaxies are part of the system. While the quasar host appears highly enriched and obscured, with AGN-like photoionization conditions, the western companion shows minimal dust extinction, low metallicity ($Z\sim0.4$\,\Zsun), and star-formation driven photoionization. The eastern companion shows higher extinction and metallicity ($Z\sim0.8$\,\Zsun) compared to the western companion, and it is at least partially photoionized by the nearby quasar. We do not find any indication of AGN in the companion sources. Our study shows that while the quasar host galaxy is already very massive ($M_{\rm dyn}>10^{11}$\,\Msun), it is still rapidly building up by accreting two relatively massive ($M_{\rm star}\sim 10^{10}$\,\Msun) companion sources. This dataset showcases the power of JWST in exposing the build-up of massive galaxies in the first Gyr of the Universe.}
\maketitle

\section{Introduction} 

Quasar host galaxies at $z>6$ are among the most massive galaxies known to populate the early Universe. They host black holes with masses of $M_{\rm BH}>10^8$\,\Msun{} \citep[e.g.,][]{farina22,mazzucchelli23}, and immense reservoirs of molecular gas, $M_{\rm H2}>10^{10}$\,\Msun{} \citep[e.g.,][]{walter03,venemans17,decarli22}, and their internal dynamics are suggestive of dynamical masses of $M_{\rm dyn}\gsim 10^{11}$\,\Msun{} \citep[e.g.,][]{neeleman21}. Furthermore, the presence of gargantuan amounts of dust ($M_{\rm dust}>10^7$\,\Msun) points to an already chemically enriched interstellar medium \citep[ISM; see, e.g.,][]{venemans18,shao19}, demonstrating that various stellar generations have already formed and evolved in these systems.

\begin{figure*}
\begin{center}
\includegraphics[width=0.9\columnwidth]{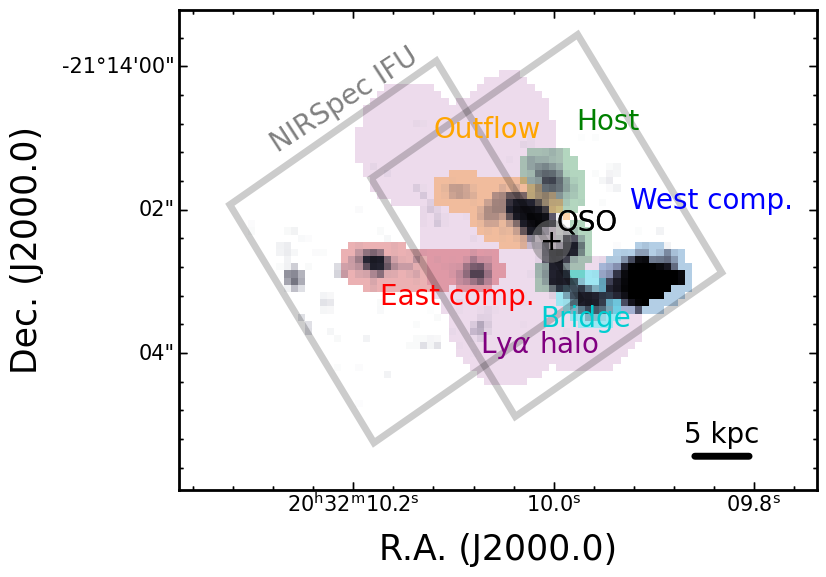}
\includegraphics[width=1.1\columnwidth]{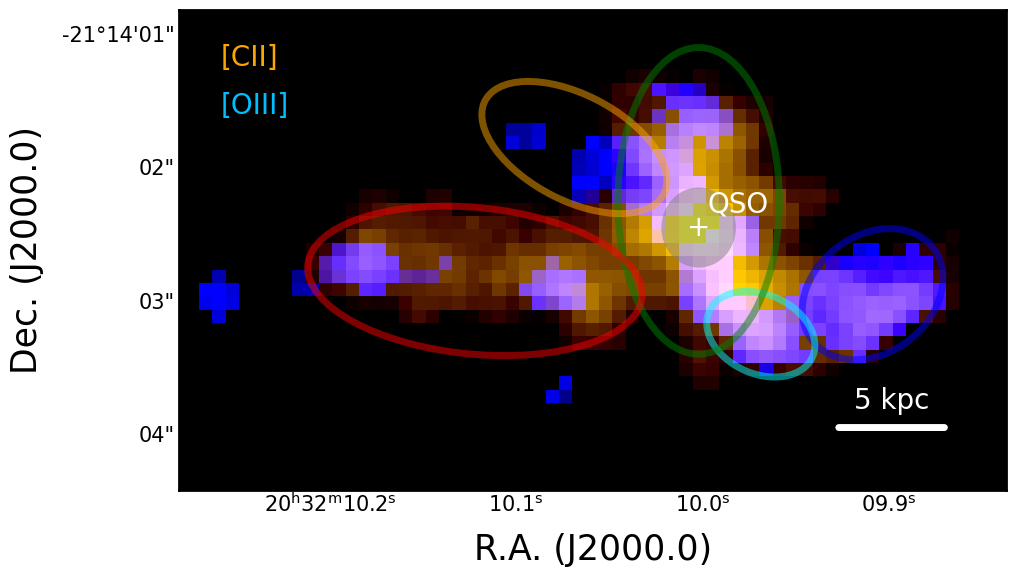}\\
\end{center}
\vspace{-4mm}
\caption{{\em Left:} The \Oiii{}$_{\rm 5007\,\AA}$ line emission map in the quasar+companion galaxies system PJ308--21 at $z$=6.2342, shown in gray scale (after PSF subtraction). The footprints of the NIRSpec IFU pointings are displayed in grey. We also mark the regions of the main components of the system with colored shading. {\em Right:} The same \Oiii{}$_{\rm 5007\,\AA{}}$ map (in blue), now with the \Cii{}$_{\rm 158\,\mu m}$ map from \citet{decarli19} superimposed. The main regions are identified with ellipses: From left to right, the eastern companion (red), the outflow (orange), the quasar host galaxy (green), the bridge (cyan), and the western companion (blue). The \Oiii{} emission arises from the regions where \Cii{} is brightest, except for the ``outflow'' component where the \Oiii{} emission stretches beyond the \Cii{} emission.}
\label{fig_system_map}
\end{figure*}

Optical/near-infrared observations (from either the ground or from space), sampling the rest-frame UV light, have so far failed to reveal the diffuse emission from gas or stars in the host galaxies of $z>6$ quasars due to a combination of intrinsic faintness, reddening, redshift, and contrast with respect to the bright nucleus \citep{mechtley12, decarli12, marshall20}. A notable exception is the ubiquitous detection of Ly$\alpha$ halos extending over several kpc around the quasars \citep[see][]{farina19}. However, Ly$\alpha$ alone can only provide limited information due to its complex radiative transfer properties. Most of what we know concerning quasar host galaxies at $z>6$ comes from observations at (sub-)mm wavelengths, which have revealed the precise redshift \citep[e.g.,][]{decarli18,eilers20}, the internal dynamics of the gas \citep[e.g.,][]{neeleman21}, as well as the content in molecular gas \citep[e.g.,][]{venemans17,li20a,decarli22}, atomic and ionized gas \citep[e.g.,][]{li20b,decarli23}, and dust \citep[e.g.,][]{venemans18,tripodi23}. A critical limitation of (sub-)mm observations is that they do not directly trace hydrogen atoms (or molecules). As a consequence, we have to rely on metallicity--dependent conversion factors and assumptions on the relative abundances of metals. Furthermore, (sub-)mm observations typically probe low-energy processes (with excitation temperatures of 10--1000\,K). Studies of the rest-frame optical emission lines can probe the Balmer series of the hydrogen atom, as well as a suite of emission lines arising from the neutral and ionized ISM (e.g., \Oiii{}$_{5007\,\AA}$, \Oiii{}$_{4363\,\AA}$, \Oii{}$_{3727\,\AA}$, \Oi{}$_{6300\,\AA}$, \Nii{}$_{6584\,\AA}$, \Sii{}$_{6717\,\AA}$, \Sii{}$_{6731\,\AA}$). Because of their accessibility in the optical band in the local Universe, these lines have been extensively used in the literature to define key diagnostics of the physical conditions of the ISM, constraining the metallicity and abundances of elements, the photoionization conditions, the electron density and temperature, and the dust reddening \citep[e.g.][]{baldwin81,kewley02,izotov06,groves23}.

Observing these lines at $z>6$ requires sensitive observations in the mid-infrared regime. The JWST \citep{gardner23,rigby23} was specifically designed to this task. Its unprecedented sensitivity at near- and mid-infrared wavelengths, paired with its excellent angular resolution, has been revolutionizing our view on the early Universe, and of the build-up of first galaxies \citep[e.g.,][]{naidu22, donnan23, eisenstein23, finkelstein23, bunker23}. Several JWST Cycle 1 programs have targeted quasars at $z>6$ and delivered the first spectra of the quasar rest-frame optical band \citep{eilers23,yang23}, the first images of the host galaxy starlight emission \citep{ding23,harikane23,yue23}, the first maps of emission lines in quasar hosts \citep{marshall23}, as well as detailed studies of the quasar environment \citep{wang23,kashino23,greene24}.

Here we present JWST/NIRSpec Integral Field Spectroscopy of the $z=6.2342$ quasar PJ308--21. This object (right ascension, R.A.=20:32:09.99, declination, Dec.=$-21$:14:02.3) was originally discovered within the Pan--STARRS \citep{chambers16} database \citep{banados16}. Follow-up observations with the Atacama Large Millimeter Array (ALMA) revealed the presence of extended dust and \Cii{} emission stretching over $\sim25$\,kpc, revealing that the quasar host galaxy is merging with one or two companion galaxies \citep{decarli17,pensabene21}. The association of the extended emission with neighbouring galaxies was further confirmed by the detection of rest-frame UV starlight continuum in Hubble Space Telescope (HST) images \citep{decarli19}. A 150 ks Chandra observation of the system revealed that the quasar is very X-ray luminous ($L_X = 2.3\times10^{45}$\,erg\,s$^{-1}$) and shows a soft spectrum (with a power law slope $\Gamma=2.4$; see \citealt{connor19}). Intriguingly, a tentative hard X-ray emission spatially consistent with the western companion of the quasar hints to the presence of an obscured AGN in this source. If confirmed, this would be a first evidence of dual AGN in the early Universe, and might shed light on the early growth of the first massive black holes.

In the first paper of this series \citep{loiacono23}, we present the observations and data reduction, and we discuss the phenomenology of the quasar. In this paper, we discuss the insights gained from the spatially resolved emission of H$\alpha$ and other rest-frame optical lines. In the third paper \citep{farina23}, we use the JWST data to address the origin of the Ly$\alpha$ halo around PJ308--21, while in the fourth paper (Lupi et al.~in prep) we discuss the gas kinematics and the dynamics in the system in connection with zoomed-in numerical simulations of quasar host galaxies at cosmic dawn.

Throughout the paper, we assume a standard $\Lambda$CDM cosmology with $H_0$=70\,\kms{}\,Mpc$^{-1}$, $\Omega_{\rm m}=0.3$, and $\Omega_\Lambda=0.7$. Within this framework, at $z$=$6.2342$ an angle of $1''$ corresponds to 5.59 kpc, and the luminosity distance $D_{\rm L}$ is 60366\,Mpc. Unless explicitly specified, all quoted uncertainties correspond to a 68\% confidence level, upper limits are reported at 3-$\sigma$ significance, and magnitudes refer to the AB photometric system \citep{oke83}.

\section{Observations and data reduction}\label{sec_observations}

The observations analyzed here were carried out on September 23, 2022 within the JWST Cycle 1 General Observer program 1554. We refer to \citet{loiacono23} for details on the observations and data processing, which we briefly summarize here. 

The observations consist of three pointings, two encompassing the PJ308--21 system (West, centered at R.A.: 20:32:09.96 and Dec.: --21:14:02.4; and East, centered at R.A.: 20:32:10.11 and Dec.: --21:14:02.8; see Fig.~\ref{fig_system_map}) and one offset by $\approx 5''$, used for background estimates. We adopted a strategy of 13 groups per integration, 1 integration per pointing, and 6 dithering positions (in a \texttt{SMALL} $\sim0.25''$ dither cycling pattern), using the \textsf{NRSIRS2} readout mode. The total integration time is 5777 sec per pointing, and each pointing encompasses an area of $\sim 3''\times3''$. We used NIRSpec in Integral Field Unit mode with the G395H grating and the F290LP filter. This provides us with spectral coverage in the range $2.87\,\mu{\rm m}<\lambda<4.03\,\mu{\rm m}$ and $4.19\,\mu{\rm m}<\lambda<5.27\,\mu{\rm m}$ at a spectral resolution of $\lambda / \Delta \lambda \approx 2700$.

We base our data reduction on our custom adaptation of the official JWST pipeline, described in detail in \citet{loiacono23}. Briefly, the JWST / NIRSpec pipeline (pipeline version: 1.12.5, context: jwst\_1183.pmap) is structured in three stages. The first one deals with detector-related calibrations (bias, dark, and cosmic rays). We apply a more severe flagging than the JWST pipeline to exclude detector ramps with too few usable steps. Stage 2 applies instrument-specific calibrations, such as flat field correction, wavelength and flux calibration. We did not perform the \textsf{ImprintStep} due to unreliable count rates in the leakage exposure. The third stage of the pipeline combines the spectra from different dither positions and creates the final data cube. Here we perform the outlier rejection as implemented in the JWST pipeline, and further apply a custom routine to further remove lingering spikes, similar to the TEMPLATES approach \citep{hutchison23}. Our aggressive flagging of the ramps and filtering of narrow spectral spikes effectively clean the final cubes of most of the cosmic ray residuals. The background in typical NIRSpec observations at these wavelengths is dominated by continuum zodiacal light and residual detector noise. We found that our observations include a sufficient number of ``source--free'' pixels that we used to assess and subtract off the background spectrum within each pointing. 

In the final stage, the JWST NIRSpec pipeline re-aligns the dithered frames and resamples the combined cube using a grid of $0.1''$ pixels. With the pipeline version in use (1.12.5), the science pointings are automatically merged in a single cube which covers a total area of $\sim 22$\,arcsec$^2$. This version of the pipeline introduces detector-related stripes in the cubes, which were not present when using previous versions of the pipeline (e.g., 1.8.3), but are visible in other JWST NIRSpec observations (Eros Vanzella, private communication), and whose origin is unclear at the moment. We treat the stripes as an additive component of the background and subtract it off from the calibrated cubes as described in Appendix \ref{sec_detector}. By comparing data cubes calibrated with different versions of the pipeline, systematic differences in the reconstructed fluxes are observed, at times as high as a few MJy\,str$^{-1}$. This translates in systematic discrepancies in the measured line fluxes that can be as high as $\sim$15\% of their value, depending on wavelength, signal-to-noise, etc. As the reliability of the calibrators and the robustness of the pipeline improve, we expect that the measured fluxes might fluctuate slightly.

The \textsf{WATA} acquisition approach adopted in the observation of PJ308--21 overrides the input astrometric barycenter. We thus correct a posteriori the astrometry of the calibrated cubes using previous HST and ALMA observations as reference. We apply a rigid shift of coordinates that places the observed position of the quasar [pixel: (47.8, 30.8)] at its nominal coordinates (R.A.: 308.0416803, Dec.: -21.2340084; J2000.0).

\section{Analysis}\label{sec_analysis}

\subsection{Point Spread Function modeling}\label{sec_psf}

The light from the quasar outshines the emission from the host galaxy at rest-frame optical wavelengths. In order to remove it, we create empirical models of the quasar Point Spread Function (PSF) using the wavelength ranges encompassing the wings of the Balmer broad emission lines \citep[see, e.g.,][]{inskip11,husemann22,marshall23}. By using the quasar light itself as reference, we circumvent any potential limitation of synthetic models (e.g., based on \textsf{WebbPSF}\cprotect\footnote{\verb|https://github.com/spacetelescope/webbpsf|}). Furthermore, this approach avoids drawbacks due to the sub-pixel alignment and intrinsic spectral dependence of the PSF models empirically derived from other astrophysical sources. Our approach requires that any extended emission is negligible compared to the quasar light in the channels used for the PSF creation. In this context, we limit the PSF analysis to the central $1''$ around the quasar, and we do not consider as reliable the residual photometry in the central $0.3''$ of the cube (marked with a circle in all relevant figures).

\begin{figure}
\begin{center}
\includegraphics[width=0.99\columnwidth]{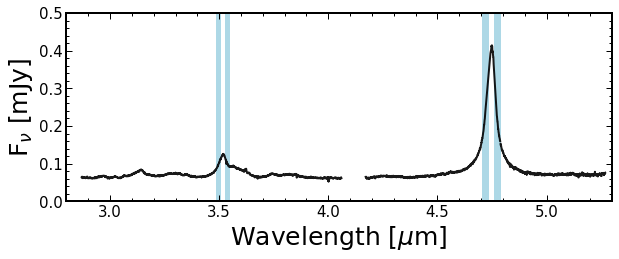}\\
\includegraphics[width=0.99\columnwidth]{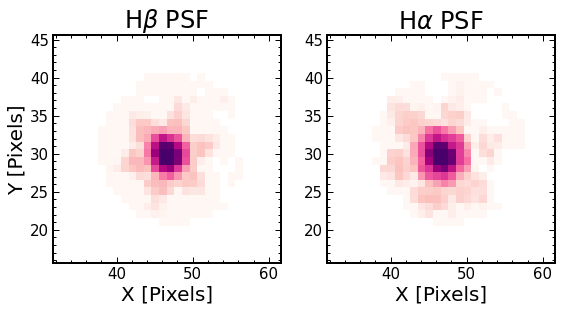}\\
\end{center}
\caption{The NIRSpec empirical Point Spread Function (PSF) models adopted in our analysis. {\em Top:} The spectrum of the targeted quasar, PJ308--21, extracted over a $0.2''$ aperture centered on the brightest pixel of the cube. The blue shaded boxes mark the channels used to model the PSF. {\em Bottom:} PSF models based on the wings of the H$\alpha$ and H$\beta$ broad emission lines.}
\label{fig_psf}
\end{figure}

We employ a custom version of the \textsc{MuseHalo} package \citep{farina19} for the PSF modeling and subtraction. We use the PSF model created by integrating over the H$\beta$ line wings for the blue part of the spectrum, and the one based on the H$\alpha$ line wings for the red part of the spectrum, in order to minimize the impact of the wavelength dependence of the PSF. We use both the blue and the red wings of the broad lines corresponding to a velocity shift of $\Delta v=\pm 800$ to $2800$\,\kms{}, or to observed-frame wavelength ranges $3.484$--$3.507$ and $3.526$--$3.550$ $\mu$m for H$\beta$, and $4.703$--$4.735$ and $4.760$--$4.792$ $\mu$m for H$\alpha$. The resulting PSF models are shown in Fig.~\ref{fig_psf}. The model is then scaled on a channel-by-channel basis to match the quasar emission in the central $0.2''$ (= 2 pixels) and subtracted.

We also tested the PSF subtraction method adopted in \textsf{q3dfit} \citet{rupke23}. In this case, the quasar spectrum is modeled, scaled off, and subtracted on a pixel-by-pixel basis. This spectrum--driven approach is best suited to recover as much of the starlight continuum as possible in regimes where the quasar is not overwhelmingly brighter than the host. However, it suffers from PSF undersampling in the bluer part of the NIRSpec spectra. This manifests itself as wiggles and sawtooth patterns which might hinder the spectral modeling \citep[see, e.g.,][]{marshall23,perna23}. Our tests show that the recovered, PSF--subtracted cubes of PJ308--21 obtained with the two approaches are largely consistent with each other. 

\subsection{Line maps, spectra, and fluxes}\label{sec_maps}

\begin{figure}
\begin{center}
\includegraphics[width=0.49\textwidth]{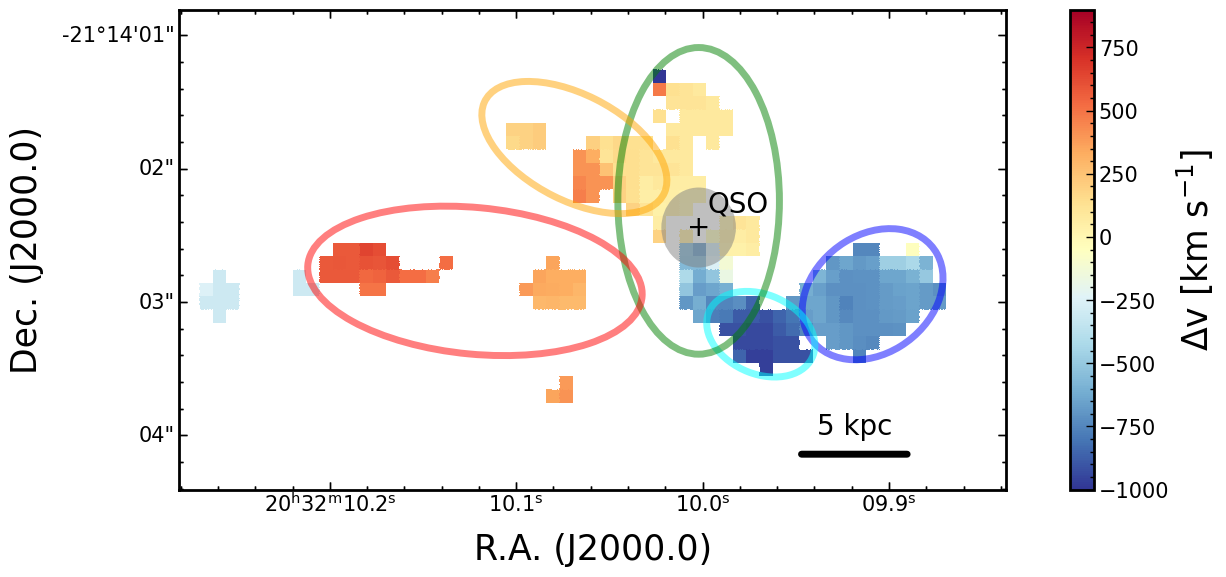}\\
\includegraphics[width=0.49\textwidth]{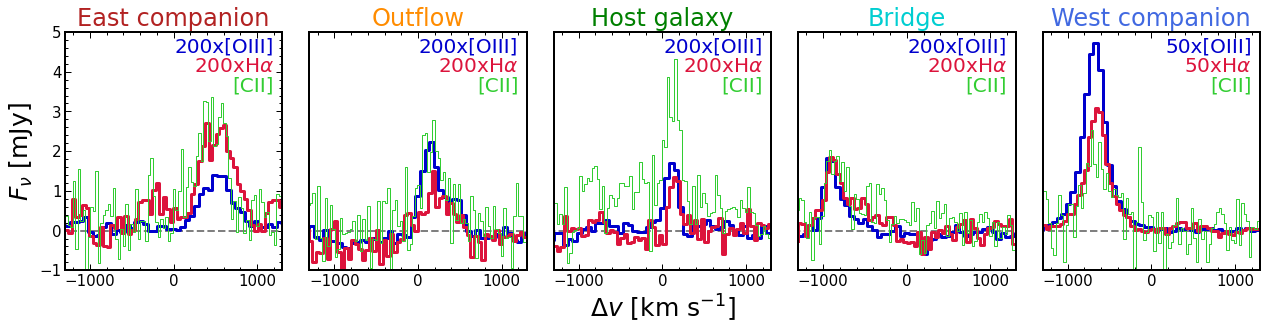}\\
\end{center}
\vspace{-4mm}
\caption{Velocity maps of \Oiii{}$_{5007\,\AA}$ in the PJ308--21 system ({\em top panel}). The system shows a prominent velocity gradient: The western companion and the bridge connecting it to the quasar host galaxies show a blue shift of $\Delta v\approx-650$  and $-880$ \kms{} respectively compared to the system rest frame ($z=6.2342$, based on the \Cii{} emission). The host galaxy shows a prominent velocity gradient from South ($\Delta v\approx -700$\,\kms{}) to North ($\Delta v = +140$\,\kms{}). The eastern companion is redshifted with respect to the system's rest frame, with $\Delta v$ increasing at increasing distance from the quasar (up to $\Delta v \approx 650$ \,\kms{}). The eastern companion shows a modest velocity dispersion ($\sigma_{\rm line}<140$\,\kms{}). }
\label{fig_gas_kin}
\end{figure}

In order to account for the large velocity gradient in the structure ($>$1000 \kms) when creating moment maps of all the relevant emission lines, we follow a two-step procedure. First, we create moment 0, 1, and 2 maps for the \Oiii{} 5007\,\AA{} line by collapsing the PSF--subtracted cube in the wavelength range $3.6055 < \lambda / {\rm [\mu m]} < 3.6361$. In this wavelength range, the \Oiii{} 5007\,\AA{} line emission is not contaminated by other significant emission lines. Fig.~\ref{fig_gas_kin} shows the velocity structure of the system, as traced by \Oiii{} moment 1 map. We then use the information from the moment 0, 1 and 2 maps to identify the voxels in the 3D cube where we expect emission from each line. Namely, in each spatial pixel, we mask out all the channels with associated wavelength $\lambda$ for which $|\lambda - \lambda_{\rm obs}\,(1+\Delta v\,c^{-1})|>3\,\lambda_{\rm obs}\,\sigma\,c^{-1}$, where $\lambda_{\rm obs}=\lambda_0 (1+z)$ is the redshifted wavelength of the line (for $z=6.2342$), $\Delta v$ and $\sigma$ are the velocity offset and velocity dispersion values as measured from the preliminary \Oiii{} moment 1 and 2 maps, and $c$ is the speed of light. We also create continuum--only images by median--averaging all the channels from the NRS1 (blue) and NRS2 (red) detectors that are not contaminated by bright emission lines (H$\beta$, \Oiii{}$_{\rm 5007\,\AA}$, H$\alpha$, \Nii{}).

\begin{figure}
\begin{center}
\includegraphics[width=0.49\textwidth]{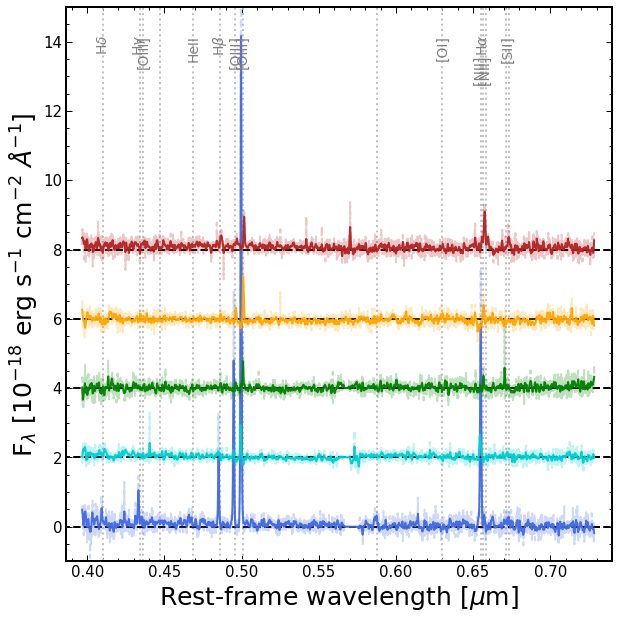}\\
\end{center}
\caption{Spectra of the main components of the PJ308--21 system, integrated over the apertures shown in Fig.~\ref{fig_system_map}, {\em left}. For the sake of clarity, spectra are shifted vertically. The spectra are ordered by increasing right ascension of the components: The west companion (blue, at the bottom), the bridge (cyan), the quasar host galaxy (green), the outflow (orange), and the east companion (red). Main emission lines are labeled in grey. The rest-frame is set to the barycenter of the system at $z=6.2342$. Thin, lighter lines show the original spectra, while thicker and darker lines show the spectra after convolution with a gaussian, with the sole purpose of increasing the readibility of the plot. The west companion shows extremely high equivalent widths ($>1000$\,\AA) in the main emission lines. A faint stellar continuum is also visible for both the east and the west companions.}
\label{fig_all_spec}
\end{figure}

The system shows a complex morphology and kinematic structure (see Fig.~\ref{fig_system_map}). From the PSF--subtracted \Oiii{} map, we identify five main regions: The western companion, the quasar host galaxy, a bridge connecting them, the eastern companion, and an additional component traced in \Oiii{} but not in \Cii{}$_{\rm 158\,\mu m}$, suggesting that it is a purely ionized gas component. Because it stretches radially from the quasar location, we interpret this as an outflow. For reference, Fig.~\ref{fig_system_map} also shows the location of the Ly$\alpha$ halo reported in \citet{farina19}. For each region of the system, we extract a spectrum by integrating over all the pixels in the apertures shown in colored shading in Fig.~\ref{fig_system_map}, {\em left}. Fig.~\ref{fig_all_spec} shows the extracted spectra.

\begin{figure*}
\begin{center}
\includegraphics[width=0.49\textwidth]{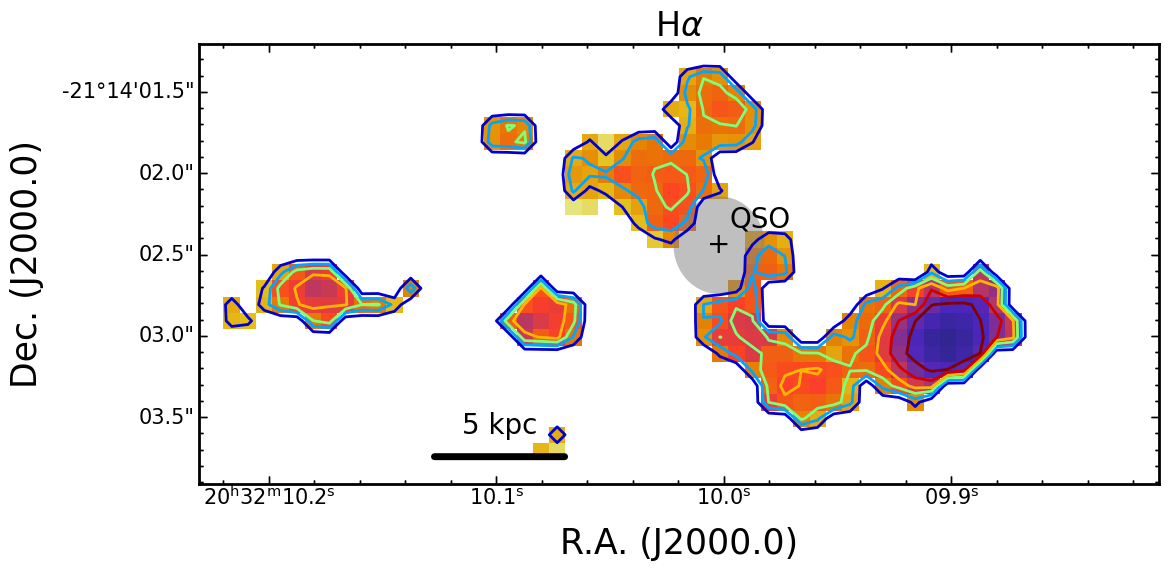}
\includegraphics[width=0.49\textwidth]{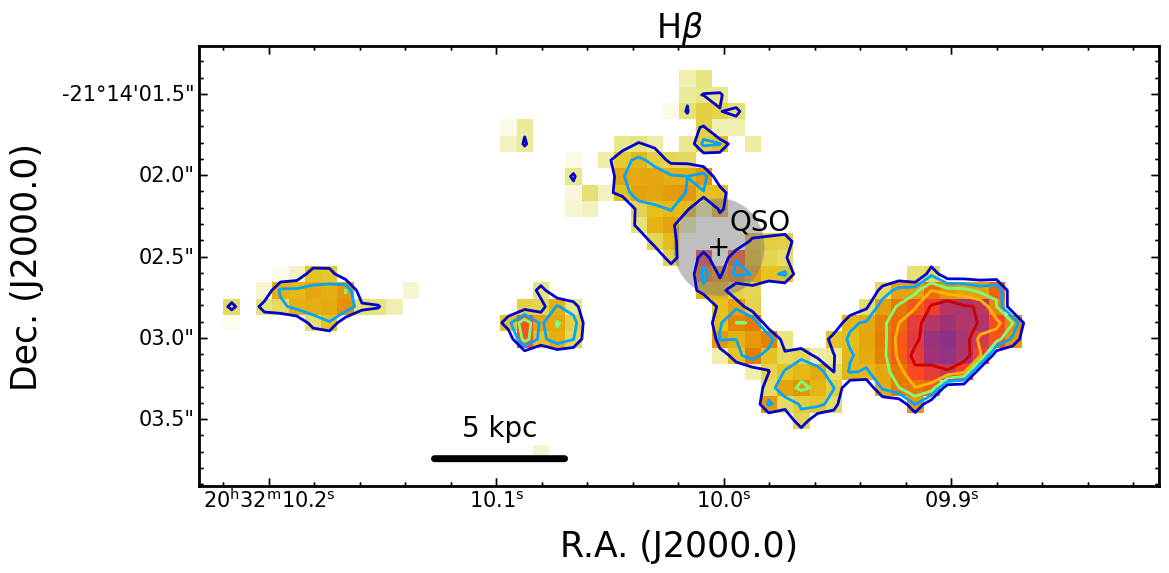}\\
\includegraphics[width=0.49\textwidth]{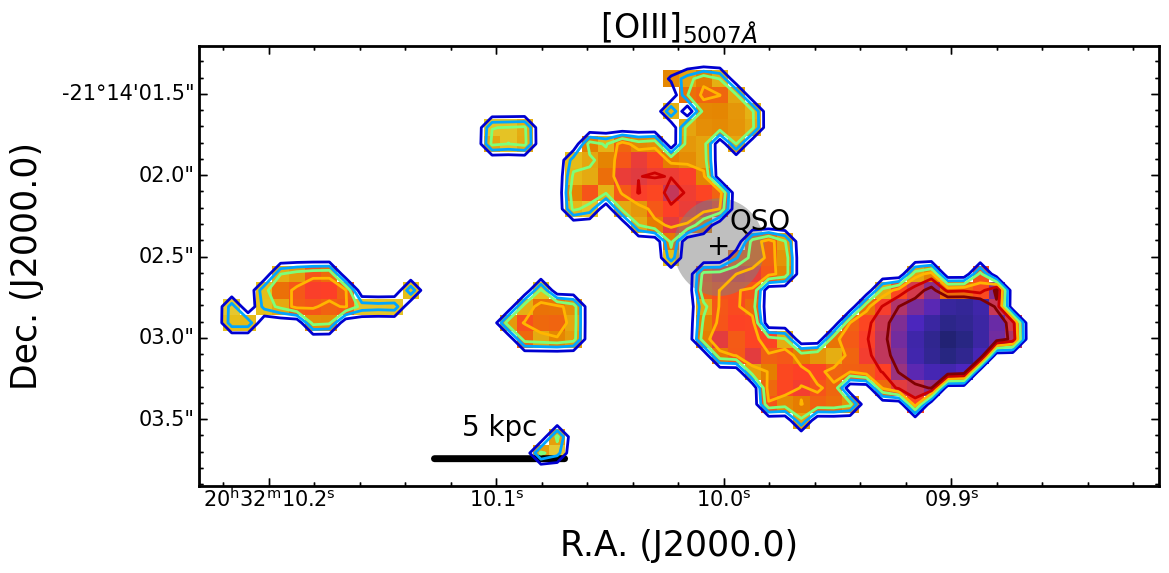}
\includegraphics[width=0.49\textwidth]{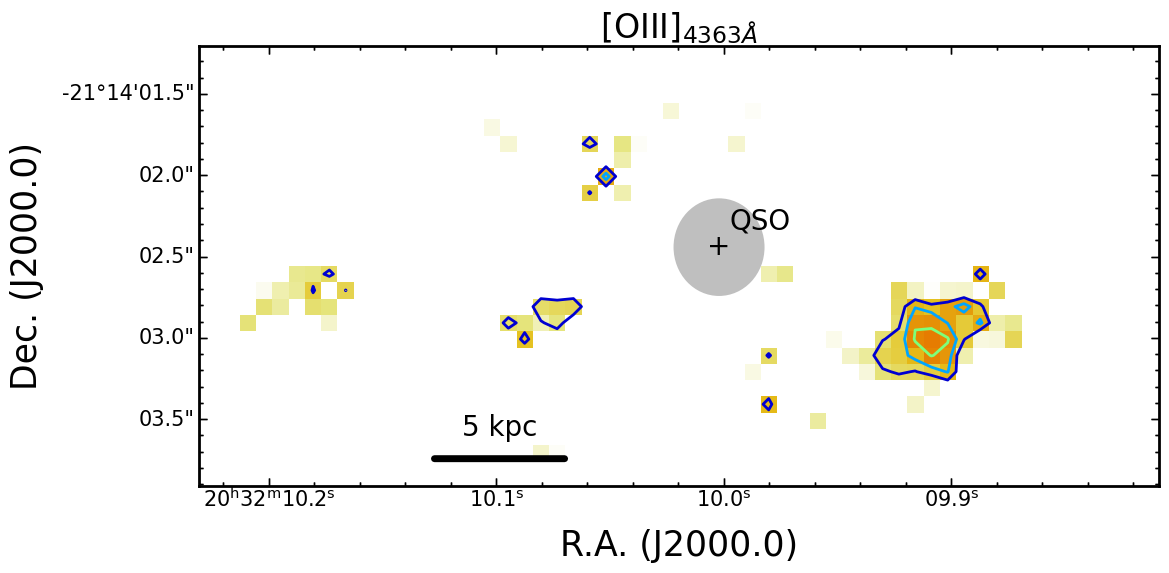}\\
\includegraphics[width=0.49\textwidth]{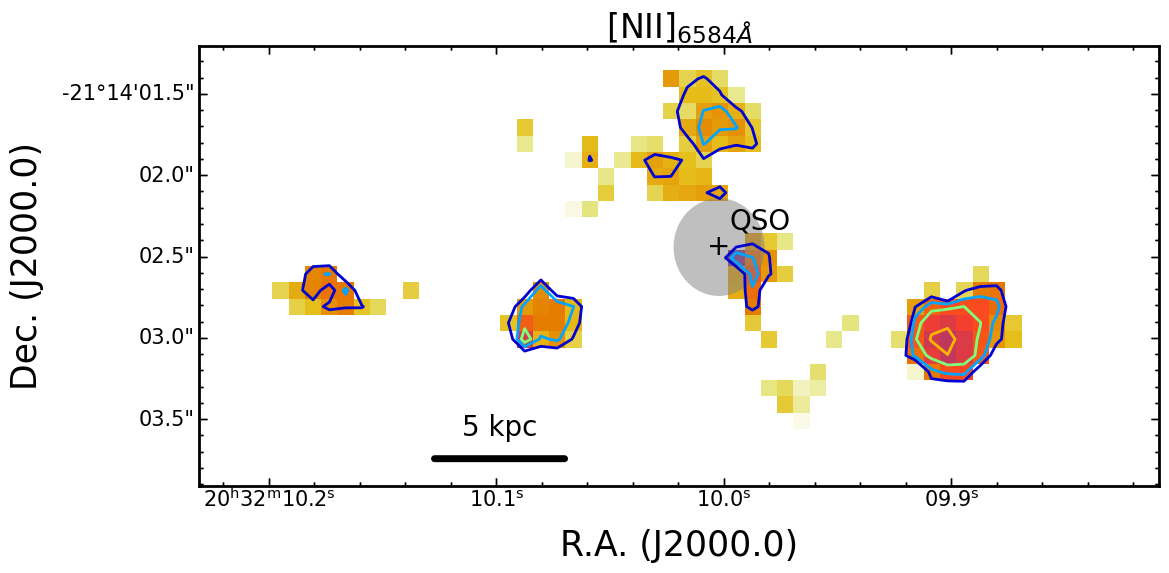}
\includegraphics[width=0.49\textwidth]{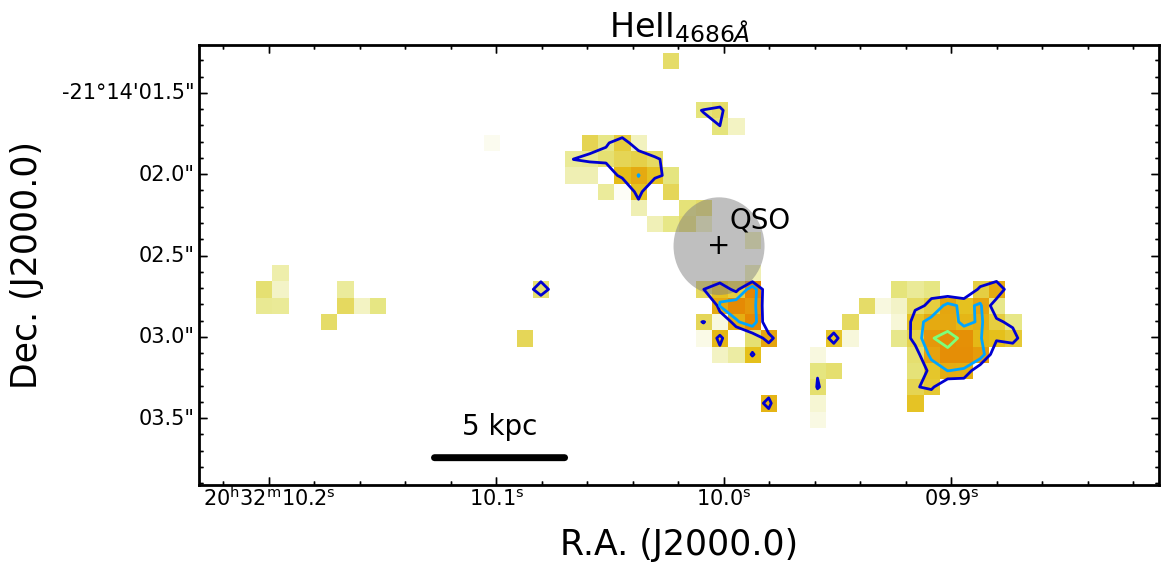}\\
\includegraphics[width=0.49\textwidth]{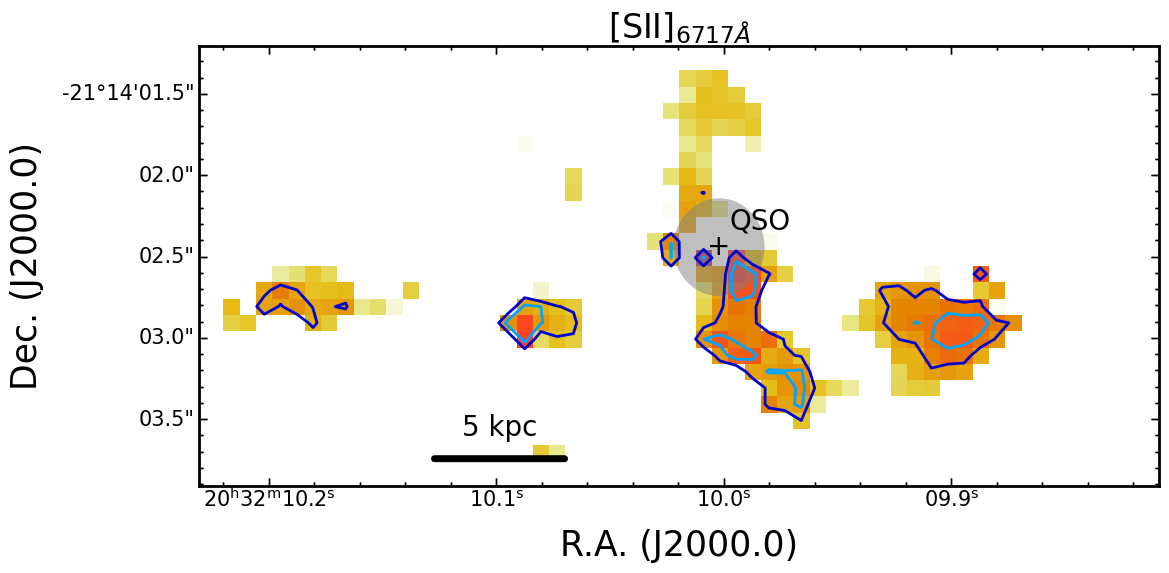}
\includegraphics[width=0.49\textwidth]{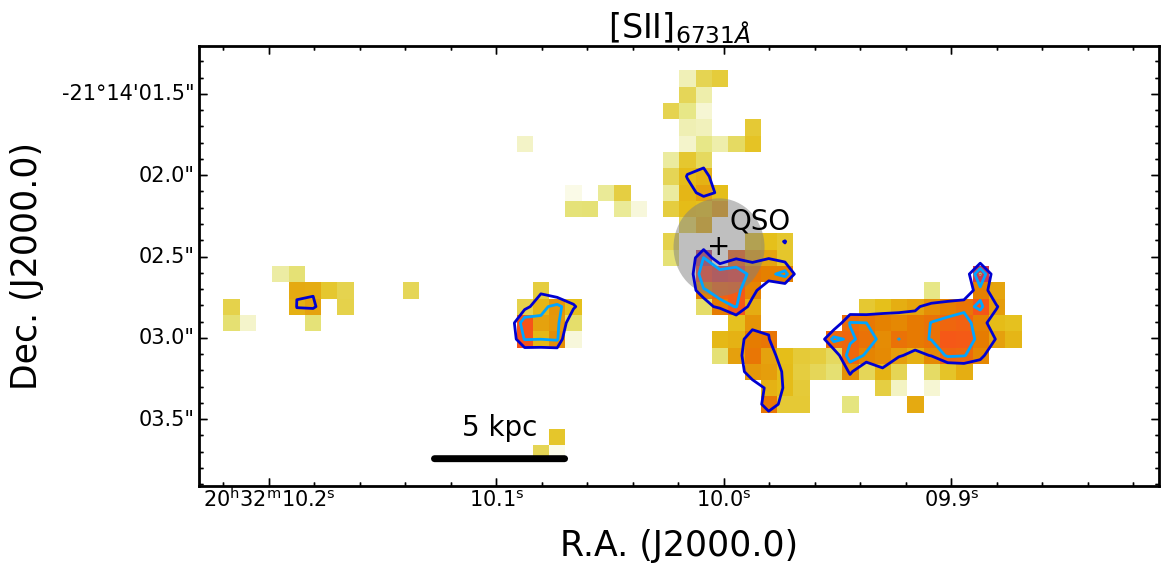}\\
\includegraphics[width=0.69\textwidth]{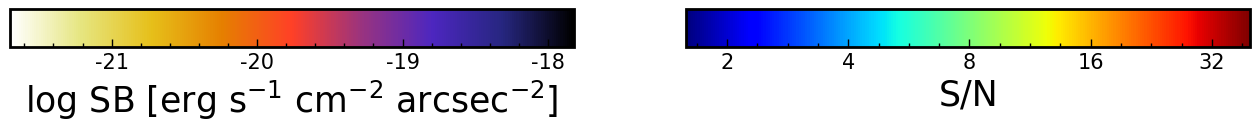}\\
\end{center}
\caption{Spatially-resolved line emission in the quasar+companion  galaxies system PJ308--21 at $z$=6.2342. The morphologies of the H$\alpha$, H$\beta$, \Oiii{}, \Nii{}, \Sii{} and \Heii{} emission are shown on the same surface brightness (SB) scale. Iso-significance contours of the line maps are shown at S/N=2, 4, 8, 16, 32.}
\label{fig_alllinemaps}
\end{figure*}

We measure line fluxes via direct integration of the spectra. First, we fit the \Oiii{}$_{\rm  5007\,\AA}$ line with a Gaussian profile, from which we infer the input line velocity shift, $\Delta v_{\rm in}$, with respect to the systemic redshift ($z=6.2342$, based on the \Cii{} emission of the quasar host galaxy) and the line width, $\sigma_{\rm in}$. Then, we measure the line flux by integrating the extracted spectra in the range corresponding to $\Delta v_{\rm in}\pm 3\times \sigma_{\rm in}$. We infer the uncertainty in the line flux by combining the nominal uncertainty of each spectral pixel within the integration window. We also refine the estimate of the velocity shift of each line component based on the flux--weighted wavelength average, $\langle \lambda \rangle = \int \lambda F_\lambda d\lambda / \int F_\lambda d\lambda $, so that the velocity shift with respect to the system redshift is $\Delta v = c\, (\langle \lambda \rangle/\lambda_{\rm obs} - 1)$. The direct integration method adopted here relies on the assumption that all of the emission lines associated with a given region map share similar kinematics. We verify that all of the brightest emission lines match this criterion. As an example, in Fig.~\ref{fig_gas_kin}, {\it bottom}, we show the line profiles of \Oiii{}$_{\rm 5007\,\AA}$, H$\alpha$ and \Cii$_{\rm 158\,\mu m}$ in each of these regions. All the observed spectral line profiles appear consistent with each other.

Tab.~\ref{tab_spec_fits} lists the results of the spectral line measurements. We consider as detected all the lines with a flux measurement larger than the upper 3-$\sigma$ confidence level.

\subsection{Radiative transfer models}\label{sec_rt}

We use radiative transfer models to interpret the observed emission line ratios in terms of underlying physical quantities. To this goal, we expand on the analysis already presented in \citet{pensabene21}, \citet{meyer22}, and \citet{decarli23}, and briefly summarized here. We take advantage of the radiative transfer code \textsf{Cloudy} \citep{ferland98,ferland13,ferland17}. 

We assume that the gas is organized in homogeneous plane-parallel slabs photoilluminated by the radiation field emitted from either young stars or an AGN. In the former case, we assume a black body spectrum with $T=50,000$\,K; in the latter case, we adopt the default AGN template from \textsf{Cloudy}. The gas density, $n$, ranges from 1\,cm$^{-3}$ to $10^6$\,cm$^{-3}$ in steps of 0.25\,dex. We also consider a range of values for the ionization parameter, $U$=$Q({\rm H})$\,$(4\pi\,r^2\,n\,c)^{-1}$, where $Q({\rm H})$ is the number of hydrogen--ionizing photons emitted per unit time, $r$ is the distance between the source of ionizing photons and the gas cloud, $c$ is the speed of light, and $n$ is the gas density. In our analysis, we consider log\,$U$ between $-4$ and 0, in steps of 0.2\,dex. Finally, following \citet{decarli23}, we adopt metallicity--dependent abundances according to the analytical prescriptions in \citet{nicholls17}. We consider a metallicity range between $10^{-1}$ and $10^{0.4}$\,Z$_\odot$ in steps of 0.1\,dex. We also include \textsf{ISM} and \textsf{PAH} grains for absorption and scattering by dust, a default background of cosmic rays, and the background radiation field from the Cosmic Microwave Background at the quasar's redshift.

\section{Results}\label{sec_results}


\subsection{Morphology and kinematics of the ionized gas}\label{sec_morphology}

We show the moment 0 maps of hydrogen H$\alpha$ and  H$\beta$, of \Oiii{}$_{\rm 5007\,\AA}$, \Oiii{}$_{\rm 4363\,\AA}$, \Nii{}$_{\rm 6583\,\AA}$, \Sii{}$_{\rm 6717\,\AA}$, \Sii{}$_{\rm 6731\,\AA}$, and \Heii{}$_{\rm 4686\,\AA}$ in PJ308--21 in Fig.~\ref{fig_alllinemaps}. To first order, all the emission lines display similar morphologies, although the relative strengths and thus the impact of the limiting surface brightness in our observations vary throughout the system.
The H$\alpha$, H$\beta$, and \Oiii{}$_{\rm 5007\,\AA}$ lines are well detected everywhere in the system. The \Oiii$_{\rm 4363\,\AA}$ is significantly detected only in the western companion. The \Nii{} emission appears distributed throughout the system, although it appears more patchy than other emission lines due to surface brightness limitations. The \Heii{} line is detected mostly in the western companion, in the quasar host galaxy, and in the outflow.
The \Sii{} emission is prominent in the bridge, and present throughout the system, although it is relatively fainter in the western companion.  These qualitative differences immediately point to different photoionization and metallicity regimes in different parts of the system.

Fig.~\ref{fig_system_map}, {\em right}, compares the morphology of \Oiii{} emission with the \Cii{} map reported in \citet{decarli19} at similar angular resolution. The gas distribution in the quasar host galaxy is best traced by \Cii{}, whereas the \Oiii{} map cannot be reliably reconstructed due to a combination of dust obscuration and contamination from the luminous quasar in the central $\sim 0.2''$. The western companion is traced by both \Cii{} and \Oiii{}, although the latter appears to extend further West-ward compared to the former, and notably, it is much brighter compared to the remainder of the system. The eastern companion shows two main knots, which are clearly identified both in their \Cii{} and \Oiii{} emission. The \Cii{} map also shows a more diffuse component that is not visible in \Oiii{}, likely due to its low surface brightness. Finally, bright \Oiii{} emission stretches from the quasar location in the North--East direction. This component does not appear visible in \Cii{}, and we identify it as an outflow. 

The gas kinematics inferred from the \Oiii{} line (the most luminous line in the JWST data examined here) closely resemble that of other transitions (e.g., H$\alpha$), and match well the kinematics of \Cii{}--emitting gas from our previous ALMA--based study (see Figs.~\ref{fig_system_map} and \ref{fig_gas_kin}). The host galaxy of the quasar shows a clear velocity gradient consistent with a rotating disk, with the Northern side receding and the Southern side approaching with respect to the systemic redshift. The peak-to-peak velocity difference is $\Delta v_{\rm peak}\approx 840$\,\kms{}. The emission stretches over $\approx 8.9$\,kpc (in diameter) or a radius of $R\approx 4.5$\,kpc. Assuming that the kinematics is driven by rotation in an edge-on (inclination angle $i=90^\circ$) disk, then $\Delta v_{\rm rot}\approx \Delta v_{\rm peak}/2 \sin i \approx 420$\,\kms{} and we can infer a dynamical mass of $M_{\rm dyn}=G^{-1}\,R\,(\Delta v_{\rm rot})^2 = 1.9\times10^{11}$\,\Msun{}. A lower inclination angle $i$ would increase our estimate of $M_{\rm dyn}$; and clearly our estimate of $M_{\rm dyn}$ relies on the assumption of virialization, which is not granted in such a complex system. The western companion is blueshifted (with a line of sight velocity difference of $\Delta v\approx -650$\,\kms{} with respect to the quasar's rest frame). The bridge connecting it to the quasar host actually shows an even higher blueshift ($\Delta v \approx -800$\,\kms{}). The Southern tip of the quasar host galaxy, which is blueshifted with respect to the system's barycenter, and the Eastern tip of the bridge arc spatially overlap, resulting in a blend of two spectral components and thus a larger inferred line width ($\sigma_{\rm line}\sim250$\,\kms{}). All of the components observed North and East of the quasar appear redshifted. The eastern companion galaxy shows a pronounced velocity gradient, from $\Delta v\approx +360$\,\kms{} to +650\,\kms{}. The North-West outflow is redshifted by $\Delta v\approx 115$\,\kms{}.

\subsection{Source of photoionization}\label{sec_bpt}

The BPT diagram \citep{baldwin81} provides us with an empirical method to discriminate between star formation and AGN-driven photoionization. In its original implementation, the diagram compares the observed line ratios between \Nii$_{\rm 6584\,\AA}$ and H$\alpha$ and between \Oiii{}$_{\rm 5007\,\AA}$ and H$\beta$. Because the involved lines have similar wavelengths, this diagnostics is rather insensitive to dust reddening and broad calibration issues. Galaxies tend to populate the diagram in a V shape, where the left--hand wing (low \Nii{}/H$\alpha$ ratios) is populated by star-forming galaxies, whereas the right-hand wing (high \Nii{}/H$\alpha$ ratios) consists of AGN and shocks. Similar diagnostic power is achieved by comparing the \Oiii{}/H$\beta$ line ratio to the \Sii{}$_{\rm 6717\,\AA}$+\Sii{}$_{\rm 6731\,\AA}$ to H$\alpha$ luminosity ratio. \citet{kewley01,kewley06} identified the following thresholds to discriminate between star-forming galaxies and AGN in the local Universe:
\begin{equation}\label{eq_nnhh}
\log \frac{\rm [OIII]}{\rm H\beta} = 0.61\,\left(\log \frac{\rm [NII]}{\rm H\alpha}-0.05\right)^{-1}+1.30
\end{equation}
\begin{equation}\label{eq_sshh}
\log \frac{\rm [OIII]}{\rm H\beta} = 0.72\,\left(\log \frac{\rm [SII]}{\rm H\alpha}-0.32\right)^{-1}+1.30.
\end{equation}
However, these thresholds implicitly rely on the fact that AGN in local galaxies tend to reside in massive, metal-rich galaxies. At low metallicities, \Nii{} and \Sii{} emission is suppressed, hence sources tend to move leftwards in the diagram, to the point that AGN and star-forming conditions might be indistinguishible at $Z\lsim 0.3$\,Z$_\odot$ \citep{groves06,hirschmann23}. Indeed, early JWST results have revealed a population of broad--line AGN in relatively low metallicity galaxies at $z>5$ with narrow-line emission consistent with star--forming galaxies \citep[see, e.g.,][]{harikane23,kocevski23,uebler23}. 

\begin{figure}
\begin{center}
\includegraphics[width=0.49\textwidth]{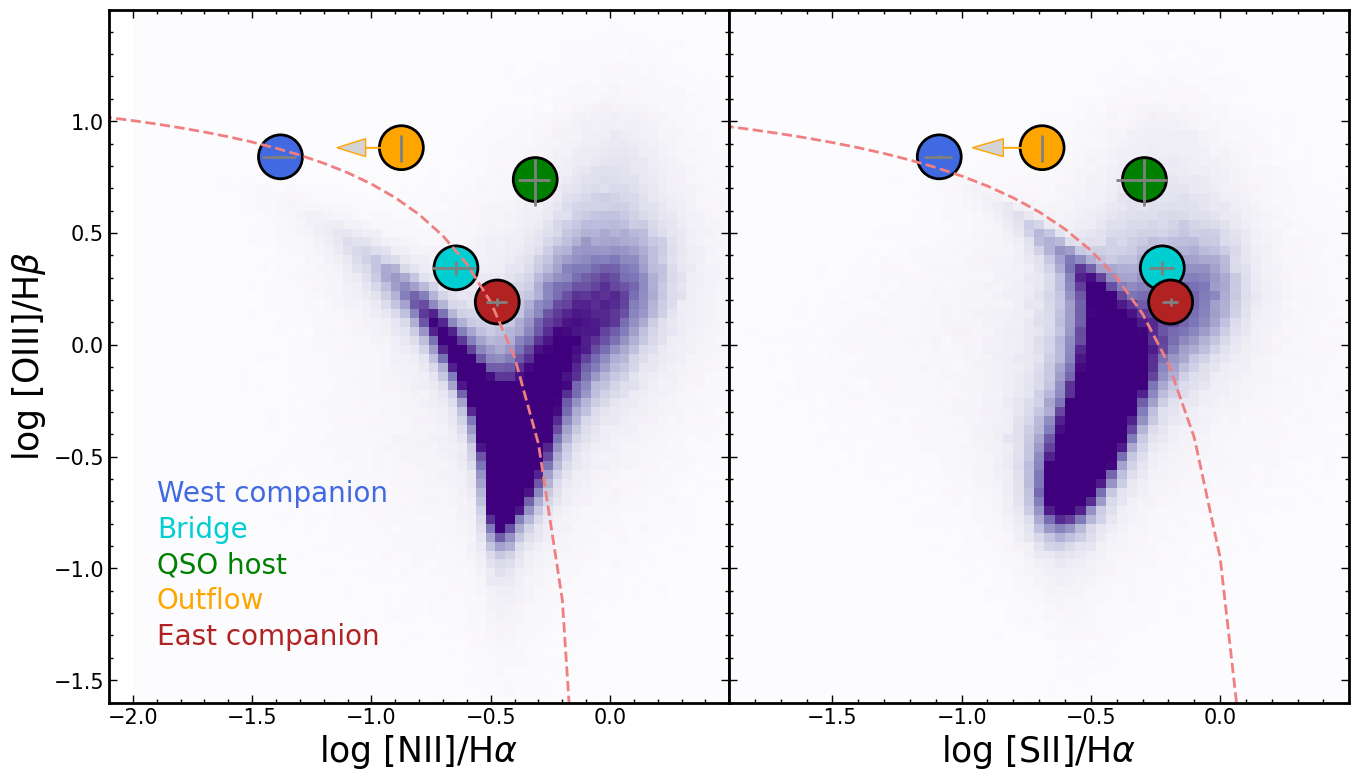}\\
\end{center}
\caption{The location of the various regions in the PJ308--21 system in the BPT diagram. The observed line ratios (and their limits) are plotted against the SDSS galaxy sample (purple shades) and the empirical thresholds between star-forming galaxies and AGN hosts (dashed lines, adopted from \citealt{kewley06}). Throughout the system, the high \Oiii{}/H$\beta$ line ratios suggest harder ionization conditions than the typical SDSS galaxy at low redshift.}
\label{fig_bpt}
\end{figure}

\begin{figure}
\begin{center}
\includegraphics[width=0.49\textwidth]{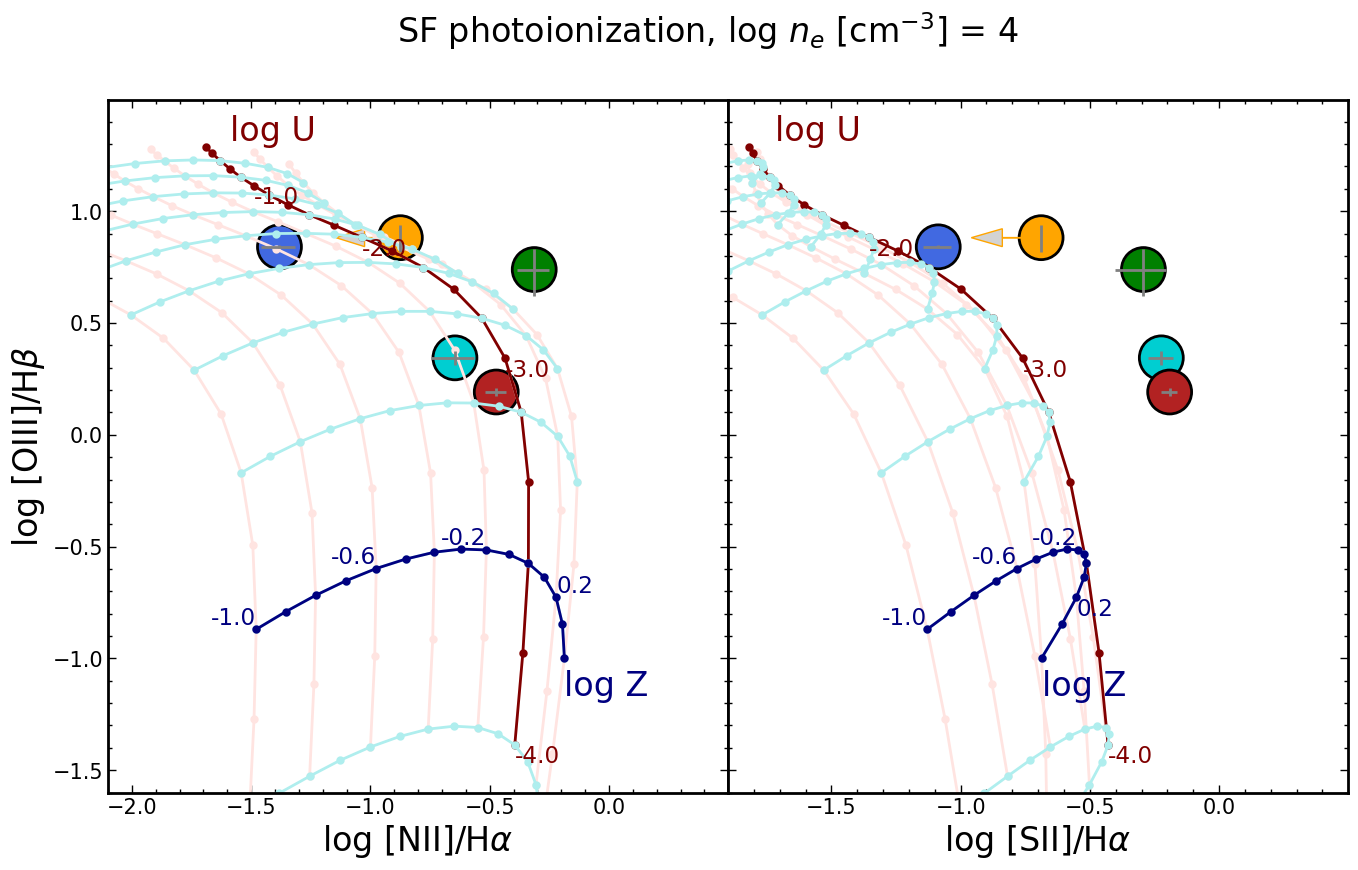}\\
\includegraphics[width=0.49\textwidth]{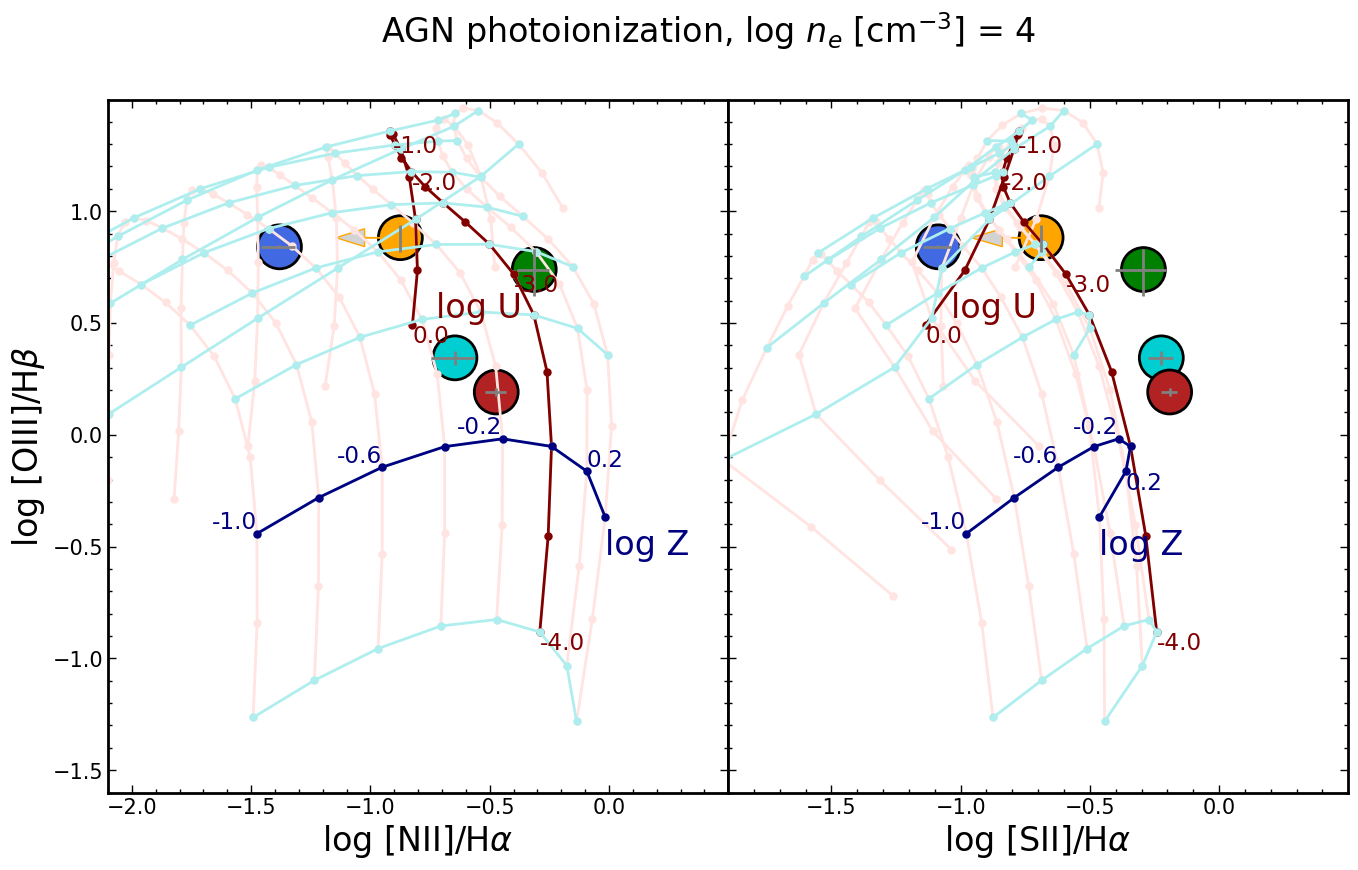}\\
\end{center}
\caption{The observed \Nii{}/\Ha{} and \Oiii{}/\Hb{} line ratios (and their upper limits)  compared to the predicted values based on our \textsf{Cloudy} models in the case of star-formation and AGN photoionization. The blue lines correspond to models with a fixed ionization parameter and varying metallicity. The red lines correspond to models with a fixed metallicity and varying ionization parameter values.}
\label{fig_bpt_models}
\end{figure}

\begin{figure}
\begin{center}
\includegraphics[width=0.49\textwidth]{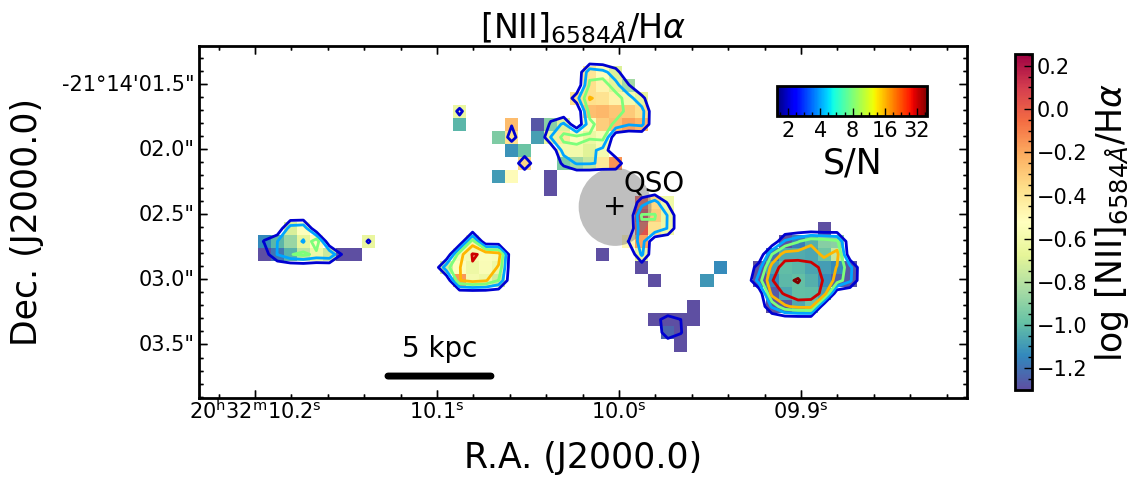}\\
\includegraphics[width=0.49\textwidth]{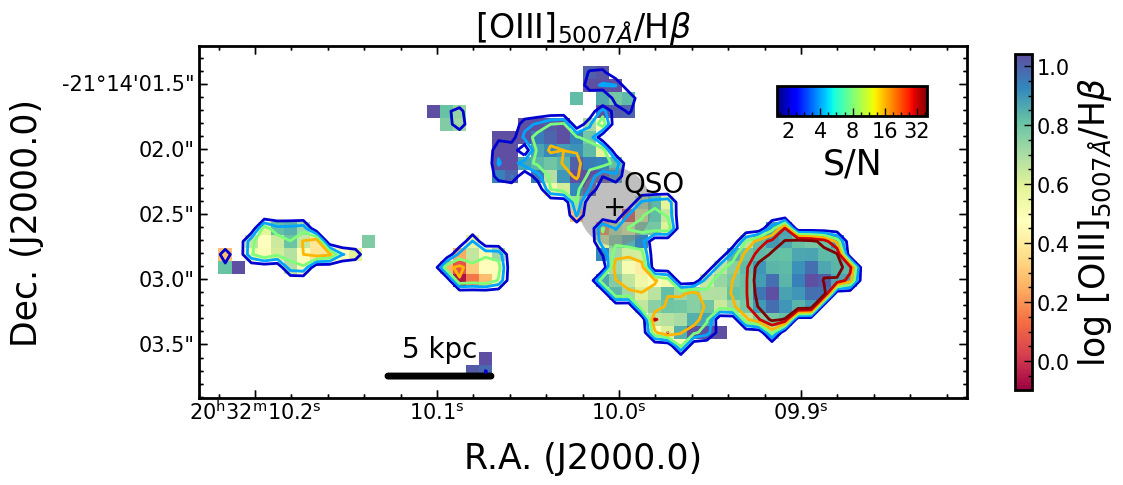}\\
\end{center}
\caption{Maps of the \Nii{}/H$\alpha$ and \Oiii{}/H$\beta$ line ratios in PJ308--21. Significance levels corresponding to S/N=2,4,8,16,32 are shown in contours. The Western companion shows low \Nii{}/\Ha{} and high \Oiii{}/\Hb{}, suggesting that its ISM is photoionized by a high-$U$ radiation field produced by young stars. The Eastern companion shows low ionization conditions consistent with star formation, although the \Nii{}/\Ha{} ratio increases at decreasing distance from the quasar -- suggesting that the quasar might contribute to the photoionization budget or that a metallicity gradient is in place. The quasar host shows AGN-like photoionization. }
\label{fig_map_source}
\end{figure}

In Fig.~\ref{fig_bpt}, we compare the observed line ratios for all the regions in the PJ308--21 system with the observed SDSS galaxy sample\cprotect\footnote{\verb|http://data.sdss3.org/sas/dr8/common/sdss-spectro/| \verb|redux/galSpecLine-dr8.fits|}, whereas Fig.~\ref{fig_bpt_models} compares the observed line ratios in the PJ308--21 system with the predictions from our \textsc{Cloudy} models
for a cloud volume density of $n=10^4$\,cm$^{-3}$ (at lower densities, to first order, all the observed line ratios decrease, as expected given that the forbidden lines in the numerators are all excited via collisions). In Fig.~\ref{fig_map_source}, we show the maps of the \Nii{}/H$\alpha$ and \Oiii{}/H$\beta$ ratios in our target on a pixel--by--pixel basis. 

While the typical SDSS galaxy can have \Oiii{}/H$\beta$ line ratios as low as 0.2, all of the components of the PJ308--21 system show high values, $>2$, with the only exception of the eastern companion, for which we measure \Oiii{}/\Hb{}$\approx$1.6. This is consistent with recent findings from other galaxy studies carried out with JWST/NIRSpec at $z>6$ \citep[e.g.,][]{sanders23,reddy23b,maseda23}. The western companion shows very low \Nii{}- and \Sii{}-to-H$\alpha$ luminosity ratios ($<0.1$), suggesting that star formation is likely the dominant photoionization process. This is further supported by the extreme equivalent width of the emission lines: e.g., the rest-frame equivalent width of \Oiii{}$_{\rm 5007\,\AA}$ is $\sim 1500$\,\AA{} in the western companion. Similar values have been reported for extreme \Oiii{} emitters and Ly-continuum leaker candidates \citep[see, e.g.,][]{izotov18}, whereas type II AGN host galaxies typically show equivalent widths $\lsim 100$\,\AA{} \citep[see, e.g.,][]{caccianiga11}. This finding challenges the robustness of the X-ray detection in the western companion \citep{connor19}. The bridge, the quasar host galaxy, and the outflow show \Oiii{}/\Hb{} ratios $>2$. The outflow and the bridge show line ratios that are consistent with both star formation and AGN as sources of photoionization. The quasar host galaxy shows relatively high \Nii{}/\Ha{} and \Sii{}/\Ha{} ratios, and \Oiii{}/\Hb{}=5.5, suggesting that its photoionization budget is mostly set by the AGN. Finally, the eastern companion appears to have a lower \Oiii{}/\Hb{} ratio than the remainder of the system, and relatively high \Nii{}/\Ha{} and \Sii{}/\Ha{} ratios, thus placing it in the locus of AGN/shocks with respect to the SDSS sample. The eastern companion shows a prominent East-to-West gradient in both line ratios, with \Nii{}/\Ha{} increasing and \Oiii{}/\Hb{} decreasing at decreasing distance from the quasar (see Fig.~\ref{fig_map_source}). This may indicate that the central quasar of PJ308--21 contributes to the photoionization budget of the eastern companion, or that this galaxy displays a metallicity gradient.

Fig.~\ref{fig_bpt_models} shows that the western companion is consistent with a star formation photoionization scenario, with $Z\lsim 0.4$\,Z$_\odot$ and $\log U\approx -2.1$. An AGN photoionization scenario would require a very low metallicity, $Z\lsim 0.2$\,Z$_\odot$ for this source, which seems in tension with our estimates (see Sec.~\ref{sec_Z}). For the eastern companion, our \textsf{Cloudy} models imply a solar metallicity $Z\sim$ Z$_\odot$, and a modest ionization parameter, $\log U \approx -3.2$, or a slightly subsolar metallicty and low values of $\log U\approx -3.5$ in the AGN scenario. The bridge, the quasar host, and the outflow are best described with AGN-like photoionization conditions, relatively high metallicities ($Z\sim$ Z$_\odot$) and a ionization parameter of $\log U \sim -2.6$.

\begin{figure}
\begin{center}
\includegraphics[width=0.39\textwidth]{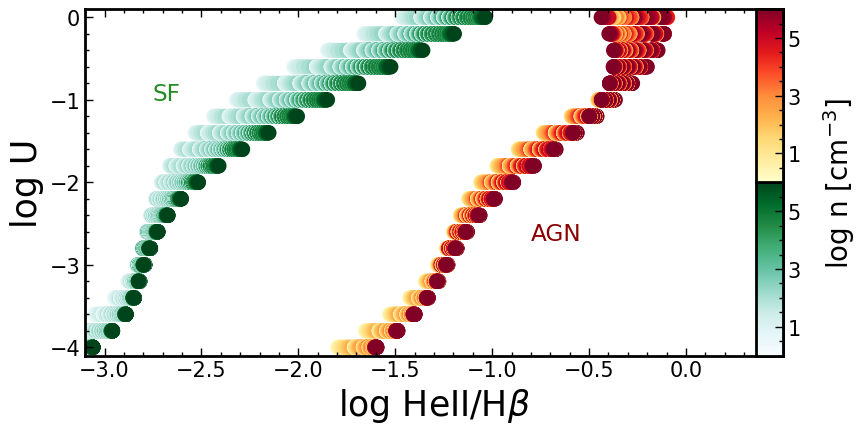}\\
\includegraphics[width=0.49\textwidth]{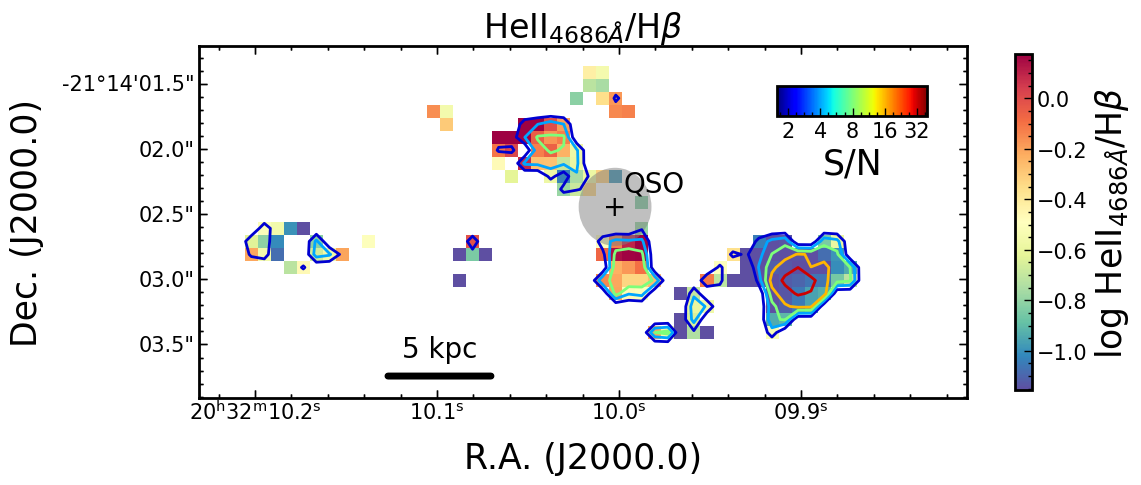}\\
\end{center}
\caption{\emph{Top:} \textsf{Cloudy} predictions for the \Heii{}/\Hb{} ratio as a function of the ionization parameter, $U$, and the gas density, $n$, for the cases of photoionization from star formation (SF, in green) and from AGN (in red). Photoionization from star formation yields \Heii/\Hb{} values $<0.1$ in all the cases. \emph{Bottom:} Map of the \Heii{}/\Hb{} line ratio in PJ308--21.  Significance levels corresponding to S/N=2,4,8,16,32 are shown in contours. Values as high as \Heii/\Hb{}$>0.5$, as found in the quasar host, in the bridge, and in the outflow are strongly indicative of photoionization from AGN. Conversely, the two companions are consistent with photoionization from star formation.}
\label{fig_map_He2Hb}
\end{figure}

\subsection{Helium photoionization}\label{sec_helium}

The diagnostics discussed in Sec.~\ref{sec_bpt} rely on the combination of metal and hydrogen lines, and thus are sensitive to abundances. Recent theoretical work \citep[based on \textsf{Cloudy} modeling; see, e.g.,][]{hirschmann23} and observational studies \citep[e.g.,][]{tozzi23} suggest that these diagnostics may not be as powerful at high redshift as in the local Universe, as galaxies might be metal poor. The He{\sc ii}/H$\beta$ luminosity ratio provides an alternative diagnostic. The He{\sc ii} line arises from recombination of He{\sc iii} ions, which is photoionized by photons with energies above 54.418 eV. Thus, the He{\sc ii}/H$\beta$ ratio is sensitive to the hardness of the radiation field. In particular, AGN photoionization results in He{\sc ii}/H$\beta$ ratios $>10$ times higher than star-formation driven photoionization for the same ionization parameter (see Fig.~\ref{fig_map_He2Hb}, \emph{top}). Once again, thanks to the modest wavelength difference between these transitions, the ratio is rather insensitive to dust reddening.

In Fig.~\ref{fig_map_He2Hb}, \emph{bottom}, we plot the spatial distribution of the He{\sc ii}/H$\beta$ ratio in PJ308--21. Our \textsf{Cloudy} models suggest that values $>$0.5 are suggestive of AGN--like photoionization, with a modest dependency on the gas density, $n$, and a strong dependence on the ionization parameter $U$. The observed ratio is $\lsim 0.1$ for the western companion. For such a low value to be consistent with AGN photoionization, the ionization parameter needs to be low, which would be in tension with the results from the analysis presented in Sec.~\ref{sec_bpt}. Thus, a star-formation driven scenario is favored. Conversely, the quasar host galaxy and the outflow show values of \Heii{}/\Hb{} of 0.5--1.0, clearly pointing to a prominent contribution from the quasar to the photoionization budget in these regions. We cannot set any meaningful constraints on the \Heii{}/\Hb{} ratio in the bridge and in the eastern companion due to their low S/N measurements.

All together, these results strongly disfavor an AGN scenario for the western companion, at odds with the tentative detection of hard X-ray photons reported in \citet{connor19}, which thus may be due to the unfortunate alignment of random background fluctuations. For the eastern companion, we have indication of a role of the central quasar in setting the photoionization conditions. Finally, the quasar host galaxy, as expected, is dominated by quasar photoionization. 


\begin{figure}
\begin{center}
\includegraphics[width=0.49\textwidth]{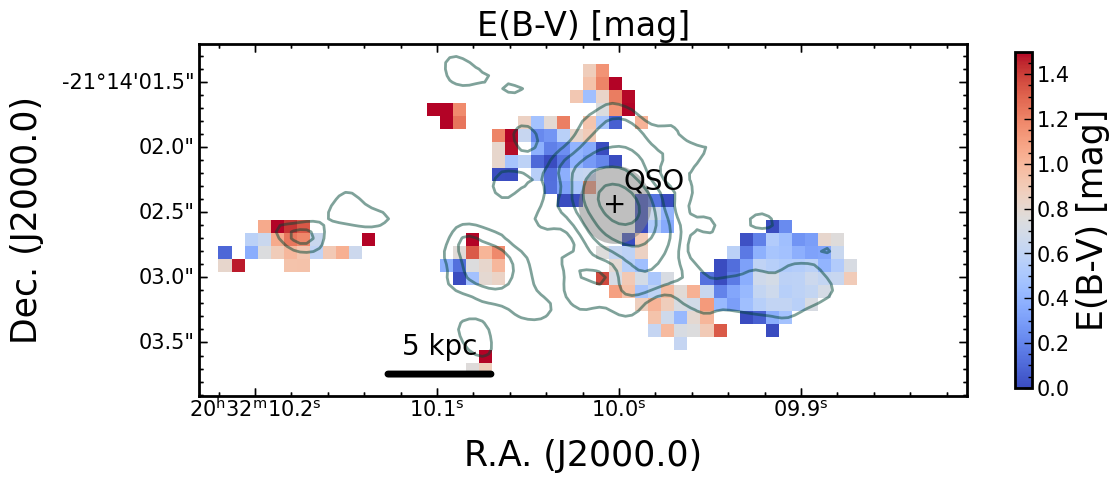}\\
\end{center}
\caption{Map of the color excess, $E(B-V)$, computed based on observed \Ha{}/\Hb{} line ratio (color scale). The western companion and the outflow show modest dust reddening, consistent with the lack of significant dust emission observed at mm wavelengths (shown in contours: log$F_\nu$/[mJy] $= -1.6, -1.5,$... $-3$; based on \citealt{decarli19}). Non negligible extinction is reported for the eastern companion and for the bridge, while very high extinction is found on the northern edge of the quasar host galaxy.}
\label{fig_map_E}
\end{figure}

\subsection{Dust reddening}\label{sec_balmerdec}

We estimate the dust reddening along the line of sight via the Balmer decrement. For standard case B recombination, assuming electron density $n_e=100$\,cm$^{-3}$ and temperature $T=10000$\,K, the intrinsic H$\alpha$/H$\beta$ flux ratio is ${\rm (H\alpha/H\beta)_{int}}$=2.86 \citep[e.g.,][]{osterbrock06}. Dust reddening increases the observed ratio. We estimate the color excess, $E(B-V)$, as:
\begin{equation}\label{eq_ebv}
E(B-V)=\frac{2.5}{\kappa(\lambda_{\rm H\beta})-\kappa(\lambda_{\rm H\alpha})}\,\log\left(\frac{\rm (H\alpha/H\beta)_{obs}}{\rm (H\alpha/H\beta)_{int}}\right).
\end{equation}
We adopt the \citet{calzetti00} law with $R_{\rm V}=3.1$\,mag as reddening curve, $\kappa(\lambda)$. The map of color excess, in magnitudes, as derived from the observed \Ha{}/\Hb{} ratio, is shown in Fig.~\ref{fig_map_E}. The western companion and the outflow are consistent with modest dust reddening. Conversely, the bridge, the eastern companion, and the northern and southern edges of the quasar host galaxy show significant reddening, $E(B-V)>0.5$\,mag or a column density $N_{\rm H}>3\times10^{21}$\,cm$^{-2}$ \citep[using the conversions reported in][]{rachford09}. We observe a qualitative agreement between the morphology of the dust emission mapped with ALMA and the reddening map derived with our NIRSpec data, with higher obscuration associated with brighter dust continuum emission (in the quasar host and in the bright knots of the eastern companion).

As a caveat, we note that the $E(B-V)$ estimates derived from the Balmer decrement are rather sensitive to measurement and calibration uncertainties. A 10\% discrepancy in the measured H$\alpha$/H$\beta$ ratio yields a color excess difference of $\delta E(B-V)\approx 0.08$\,mag, or $\delta A_V\approx 0.25$\,mag.

\begin{figure}
\begin{center}
\includegraphics[width=0.49\textwidth]{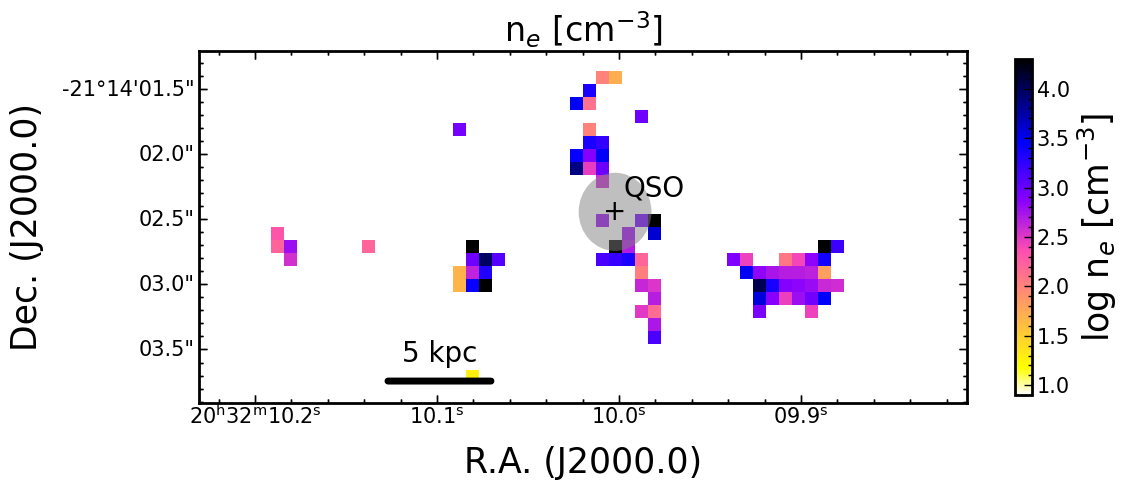}\\
\includegraphics[width=0.49\textwidth]{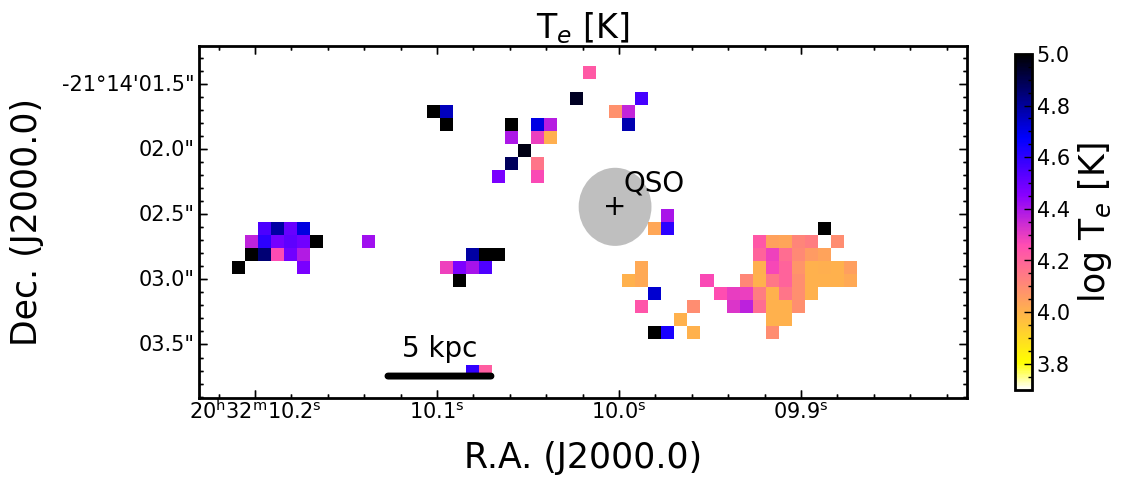}\\
\includegraphics[width=0.49\textwidth]{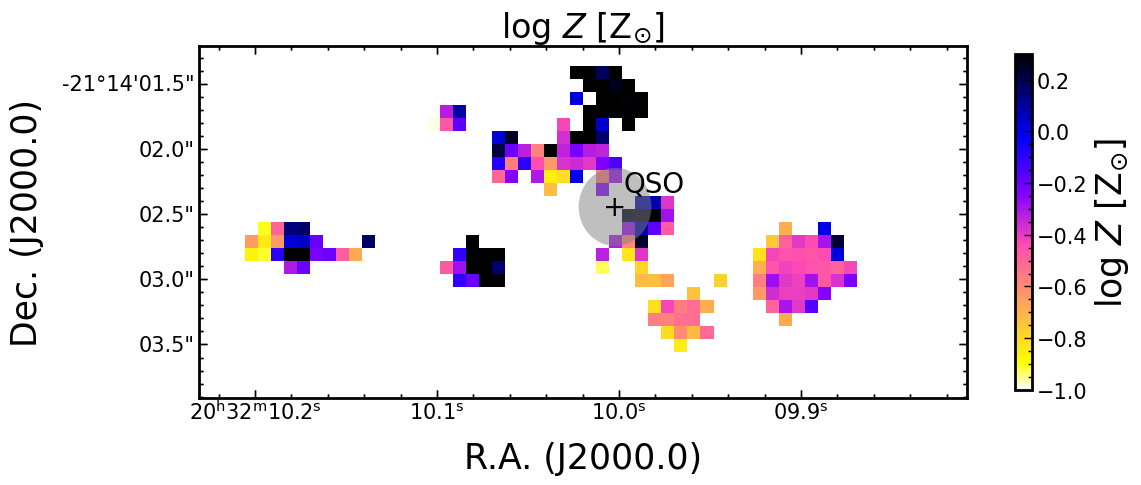}\\
\end{center}
\caption{Maps of the electron density, $n_e$ ({\rm top}), based on the \Sii{}$_{6717\,\AA{}}$/\Sii{}$_{6731\,\AA{}}$ ratio; of the electron temperature, $T_e$, based on the \Oiii{}$_{5007\,\AA{}}$/\Oiii{}$_{4363\,\AA{}}$ line ratio; and of the gas--phase metallicity $Z$, based on the Scal-PG16 method (see text for details).}
\label{fig_map_electron_nT}
\end{figure}

\subsection{Ionized gas density}\label{sec_ndens}

The ionized sulfur lines at 6717 and 6731\,\AA{} have similar excitation energy ($E/k_{\rm b}=21416$ and 21370\,K, respectively). In the low density regime ($n<10$\,cm$^{-3}$), the ratio is thus fixed, \Sii{}$_{\rm 6717\,\AA}$/\Sii$_{\rm 6731\,\AA}\sim1.5$. At higher densities, collisional de-excitation suppresses the 6717\,\AA{} transition first [critical densities: $n_{\rm crit}$(6717\,\AA)=$1570$\,cm$^{-3}$, $n_{\rm crit}$(6731\,\AA)=$14900$\,cm$^{-3}$, for a gas kinetic temperature of $10^4$\,K, assuming electrons as collision partners], thus the line ratio is sensitive to the gas density in the regime $50 < n\,{\rm [cm^{-3}]} < 50,000$. This diagnostic is rather insensitive to the source of the photoionization radiation, the ionization parameter, and metallicity (provided that the \Sii{} emission is bright enough to be significantly detected). We adopt the analytical scaling by \citet{jiang19}:
\begin{equation}\label{eq_ne_from_S2}
n_e(R) = \frac{a b - R c}{R - a}
\end{equation}
where $R=$ \Sii{}$_{6717\,\AA}$/\Sii{}$_{6731\,\AA{}}$, $a=0.4441$, $b=2514$, $c=779.3$ \citep[see also][]{liu23}. We verified that this analytical approximation is consistent within a few percent to predictions based on our \textsf{Cloudy} models.

Fig.~\ref{fig_map_electron_nT} shows the map of $n_e$ derived from the observed \Sii{} line ratio. Throughout the system, we measure electron density values around 1000\,cm$^{-3}$, although this estimate is robust only for the west companion due to sensitivity limitations. These high density values are in qualitative agreement with the findings by \citet{reddy23a,reddy23b}, who report a correlation between $n_e$ and $U$. The relatively high electron density values measured throughout the PJ308--21 system are consistent with the values reported by \citet{isobe23} for a sample of star-forming galaxies at $z=4-9$ with similar stellar mass as the companions studied here.

\subsection{Electron temperature}\label{sec_Te}

Our NIRSpec IFU observations of PJ308--21 are sensitive to the \Oiii{} 5007 and 4959\,\AA{} nebular lines, as well as to the \Oiii{} 4363\,\AA{} auroral line. We take advantage of the difference in the involved energy levels to set constraints on the kinetic temperature of electrons, $T_e$. Following eq.~5.4 in \citet{osterbrock06}, we adopt:
\begin{equation}\label{eq_O3_Te}
\frac{\rm [OIII]\,4959+5007}{\rm [OIII]\,4363} = \frac{7.90\,e^{32900/T_e}}{1+0.00045 \,n\,T_e^{-0.5}}
\end{equation}
for $T_e$ expressed in K and $n$ in units of cm$^{-3}$. This diagnostic is unaffected by gas density for $n\lsim10^5$\,cm$^{-3}$. 

In Fig.~\ref{fig_map_electron_nT}, we compare $T_e$ values computed using eq.~\ref{eq_O3_Te} with the observed values of \Oiii{} (4959+5007)/4363. The western companion shows values of $T_e=12700\pm1000$\,K, whereas constraints on the remainder of the system are uninformative. The $T_e$ value measured in the western companion is in line with what is reported for H{\sc ii} regions in local galaxies in low-metallicity ($Z<0.5$\,Z$_\odot$) environments \citep[see, e.g.,][]{ho19}.

We note that the intrinsic Balmer ratio (H$\alpha$/H$\beta$)$_{\rm int}$ adopted in Sec.~\ref{sec_balmerdec} is a function of $T_e$. While the expected ratio due to recombination changes by less than 1\% for $T_e$ increasing from 10000\,K to 12700\,K, the impact of collisional excitation of H$\alpha$ (with neutral hydrogen atoms as collisional partner) might be non-negligible \citep[see, e.g.,][]{luridiana09}. However, the correction strongly depends on the fraction of neutral gas within the nebula. The high equivalent width of the ionized gas emission lines and the high ionization parameter point to a high ionized gas fraction in the western companion, $n_p/n_{\rm HI}\ll 0.1$, yielding negligible corrections in (H$\alpha$/H$\beta$)$_{\rm int}$.

\subsection{Metallicity diagnostics}\label{sec_Z}

We base our gas-phase metallicity on the Scal-PG16 prescription \citep{pilyugin16,kreckel19,groves23}, which is based on three standard diagnostic line ratios:
\begin{equation}\label{eq_defN2}
{\rm N_2=([NII]\lambda 6548+\lambda6584)/H\beta}
\end{equation}
\begin{equation}\label{eq_defS2}
{\rm S_2=([SII]\lambda 6717+\lambda6731)/H\beta}
\end{equation}
\begin{equation}\label{eq_defR3}
{\rm R_3=([OIII]\lambda 4959+\lambda 5007)/H\beta}.
\end{equation}
The gas metallicity is then computed as:
\begin{multline}\label{eq_Z_groves}
12 + {\rm log(O/H)} = a_0 + a_1 \log(R_3 / S_2 ) + a_2 \log N_2 + \\
+[a_3 + a_4 \log(R_3 / S_2 ) + a_5 \log N_2 ] \times \log S_2
\end{multline}
with ($a_0$, $a_1$, $a_2$, $a_3$, $a_4$, $a_5$) = (8.424, 0.030, 0.751, $-0.349$, 0.182, 0.508) for $\log$ N$_2 > -0.6$, and (8.072, 0.789, 0.726, 1.069,  $-0.170$, 0.022) elsewhere \citep[see][]{groves23}. We use the extinction--corrected line measurements in eqs.~\ref{eq_defN2}--\ref{eq_defR3} based on the $E(B-V)$ estimates from the previous session.

Fig.~\ref{fig_map_electron_nT} shows the resulting map. We find that the western companion has a metallicity of $Z\sim 0.4$\,Z$_\odot$, while the eastern companion has a metallicity of $Z\sim 1$\,Z$_\odot$, consistent with the value derived based on the BPT diagram analysis via our \textsf{Cloudy} models in Sec.~\ref{sec_bpt}. Finally, the outflow shows a metallicity $Z\sim 0.5$\,Z$_\odot$.

\subsection{Stellar masses}\label{sec_mstar}

\begin{figure}
\begin{center}
\includegraphics[width=0.49\textwidth]{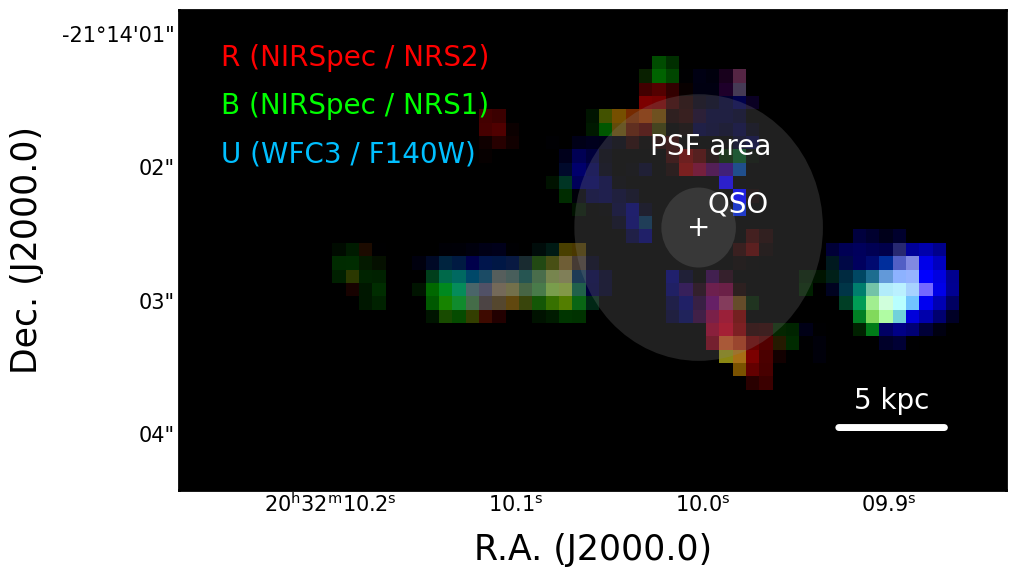}\\
\end{center}
\caption{Continuum images of the stellar populations in the rest-frame R, B, and U bands, based on the stack of the red, blue parts of the JWST/NIRSpec data (corresponding to detectors NRS2 and 1, respectively), as well as on the HST F140W image from \citet{decarli19}, resampled on the same pixel scale. The images are shown after subtracting the quasar light, and masking the regions not associated with the companions. The two galaxies are marginally detected in the JWST observations. The western companion appears bluer than the eastern companion, although the significance is modest.}
\label{fig_starlight_map}
\end{figure}

Fig.~\ref{fig_starlight_map} shows the maps of the starlight continuum, created as described in Sec.~\ref{sec_maps}, together with the HST F140W image presented in \citet{decarli19}, resampled to the same pixel scale. Both companions show some continuum emission, although their photometry is uncertain due to significant residuals in the data reduction (masked out in Fig.~\ref{fig_starlight_map}). We estimate that the western companion is detected at S/N $\sim 6.7$ (0.6) in B (R) band; whereas we detect the eastern companion at S/N $\sim 5.1$ (1.3) in B (R) band.

\begin{figure}
\begin{center}
\includegraphics[width=0.49\textwidth]{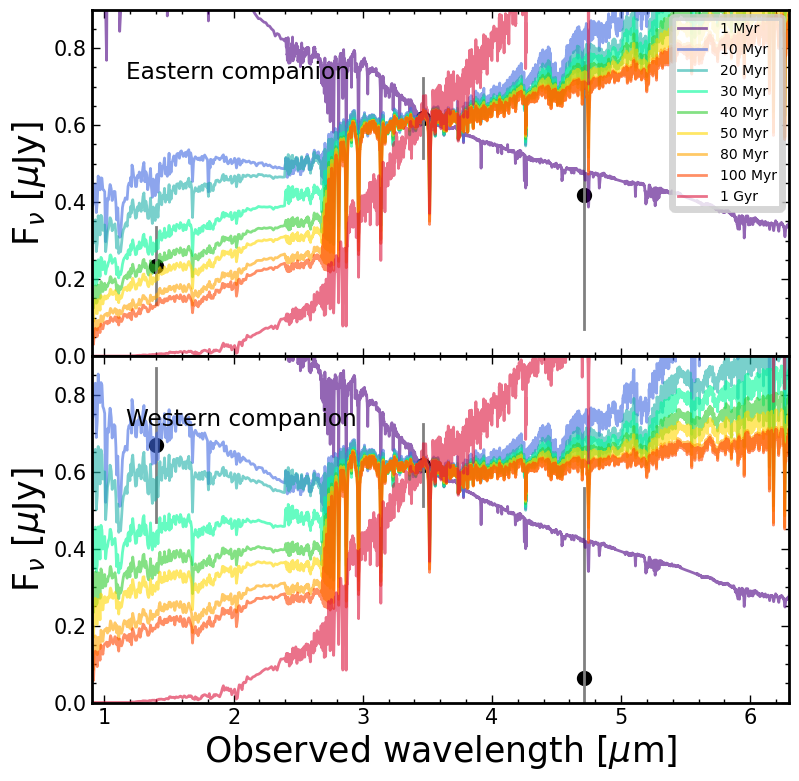}\\
\end{center}
\caption{Constraints on the starlight emission in the Eastern and Western companion galaxies, based on the HST data from \citet{decarli19} and the JWST data presented here (black circles). For comparison, single-stellar population models at various ages from \citet{bc03} are plotted, after normalizing them to the rest-frame B band data point, and after correcting for dust extinction. The observed photometry is consistent with both galaxies having a young (20--40\,Myr) stellar population and a stellar mass of $\sim (7$--$12)\times10^9$\,\Msun{}.}
\label{fig_stars}
\end{figure}

We compare the observed photometry of the starlight continua in these two galaxies with the HST--based photometry published in \citet{decarli19}, sampling the rest-frame U-band emission, and with stellar population models (Fig.~\ref{fig_stars}). The latter are single stellar population \citet{bc03} models with solar stellar metallicity and a range of burst ages, from 1 Myr to 1 Gyr, scaled to match the observed flux density in the rest-frame B band. Within the limitation of the S/N of our observations, this basic approach provides us with a rough estimate of the stellar mass of the two galaxies. 

We find that the western companion is very blue, consistent with a young ($\sim20$\,Myr) stellar population, with a stellar mass of $\sim 7\times10^9$\,\Msun{}. The eastern companion appears redder, because of both higher dust reddening and a slightly older stellar population ($\sim 40$\,Myr old). Its stellar mass is $\sim 1.4\times10^{10}$\,\Msun{}.

\subsection{Star formation rates}\label{sec_sfr}

We estimate star formation rates (SFR) from the H$\alpha$ emission, from the rest-frame UV, and from the dust emission observed at mm wavelengths. We convert the observed extinction--corrected H$\alpha$ luminosity into SFR following \citet{kennicutt12}: 
\begin{equation}\label{eq_sfrHa}
\log \frac{\rm SFR}{\rm [M_\odot\,yr^{-1}]} = \log \frac{L_{\rm H\alpha}}{\rm [erg\,s^{-1}]}-41.27.
\end{equation}
We infer a SFR(H$\alpha$) = $41.1\pm 0.3$\,\Msun{}\,yr$^{-1}$ for the western companion, and $3.5\pm0.1$\,\Msun{}\,yr$^{-1}$ for the eastern companion. 

For comparison, from the rest-frame U-band we infer SFR(UV)=12.3 and 4.8 \Msun{}\,yr$^{-1}$ for the western and eastern companions, respectively; and from the dust continuum flux density at 158\,$\mu$m, assuming a dust temperature of $T_{\rm dust}=35$\,K \citep[based on the multi-wavelength observations presented in][]{pensabene21}, we infer SFR(IR)=35.1 and 38.4\,\Msun{}\,yr$^{-1}$ for the western and eastern companions, respectively\footnote{The values of SFR(UV) and SFR(IR) reported here differ slightly from the ones reported in \citet{decarli19}, as we here consider only the emission within the masks shown in Fig.~\ref{fig_system_map} for the sake of internal consistency.}. The sum of UV- and IR-based SFR estimates yields a total SFR of 50.6 and 45.5\,\Msun{}\,yr$^{-1}$ for the western and eastern companions, respectively.

\begin{figure}
\begin{center}
\includegraphics[width=0.49\textwidth]{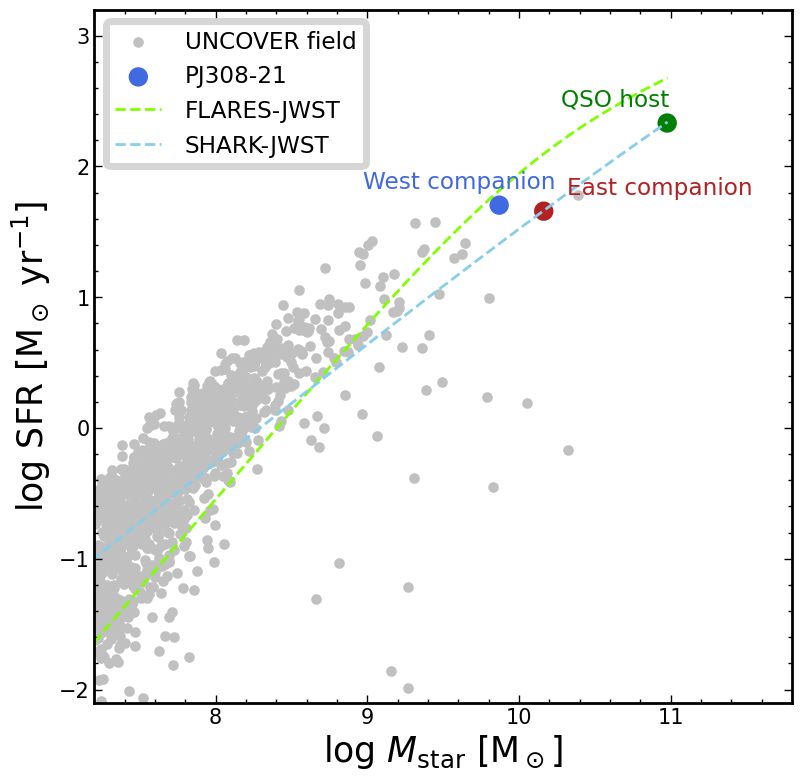}\\
\end{center}
\caption{Comparison between the stellar mass and the star formation rate estimates for the eastern (red) and western (blue) companion galaxies in PJ308--21. We also show the location of the quasar host galaxy, assuming that the stellar mass takes accounts for 50\% of the dynamical mass, and that the star formation rate is dominated by the emission traced by the dust. For comparison, the best fit main sequence of star forming galaxies at $z=6$ based on the FLARES-JWST simulations and the SHARK-JWST simulations are shown as dashed lines \citep{dsilva23}, while the sample of photometrically--selected $5.5<z<7.0$ galaxies from the UNCOVER survey \citep{uncover24} are shown as grey points. Both companion galaxies in PJ308--21 appear to lie along the expected main sequence at these redshifts, at the high--mass end of the galaxy population observed at these redshifts.}
\label{fig_MS}
\end{figure}

Fig.~\ref{fig_MS} compares the stellar masses and the total SFR estimates in these two galaxies with the expected locus of the main sequence of star-forming galaxies at similar redshifts. We refer here to the theoretical work by \citet{dsilva23}, and consider the FLARES-JWST and SHARK-JWST simulations for reference, for galaxies at $z=6$. We also compare with the sample of photometrically--selected galaxies at $5.5<z<7.0$ from the UNCOVER survey \citep{uncover24}. 

Both the companion galaxies appear to lie along the main sequence of star-forming galaxies at these redshifts if IR- or UV-based SFRs are considered. Most notably, the two companion galaxies of PJ308--21 are among the most massive galaxies in the UNCOVER sample at similar redshifts. The stellar mass function of field galaxies at $z\sim 6$ \citep[e.g.,][and references therein]{stefanon21} suggests a volume density of $\approx 6\times 10^{-4}$ galaxies per comoving Mpc$^3$ with $\log M_{\rm star}$\,[\Msun]$>$9.5. The fact that two such galaxies are in close interaction with the even more massive quasar host ($M_{\rm dyn}=1.9\times10^{11}$\,\Msun{}, see Sec.~\ref{sec_morphology}) reinforces the evidence that luminous quasars reside in prominent galactic overdensities \citep[see, e.g.][]{decarli17,mignoli20,wang23} in the early Universe.

\subsection{The star formation law}\label{sec_ks}

We use the \Cii{} map of PJ308--21 to infer the surface gas density, $\Sigma_{\rm gas}$, following, e.g., Eq.~2 in \citet{venemans17b}. We assume optically thin line emission and an excitation temperature $T_{\rm ex}=100$\,K. We adopt the analytical prescription for the carbon abundance dependence on metallicity by \citet{nicholls17}, which yields [C/H]=(3.11, 0.64, 2.23)$\times 10^{-4}$ for $Z=(1.1, 0.4, 0.9)$ (corresponding to the quasar host, the west companion, and the east companion respectively). Finally, we adopt a fiducial ionized carbon fraction of [C$^+$/C]=0.5 \citep[see, e.g.,][]{decarli23}. The star formation rate density, $\Sigma_{\rm SFR}$, is computed from dust luminosity, assuming an emissivity index $\beta=1.6$ and a dust temperature $T_{\rm dust}=35$\,K for all the components except for the quasar host galaxy, for which we adopt $T_{\rm dust}=45$\,K \citep[see][]{pensabene21}; and from the H$\alpha$ and UV luminosities, as described in Sec.~\ref{sec_sfr}. 

\begin{figure}
\begin{center}
\includegraphics[width=0.49\textwidth]{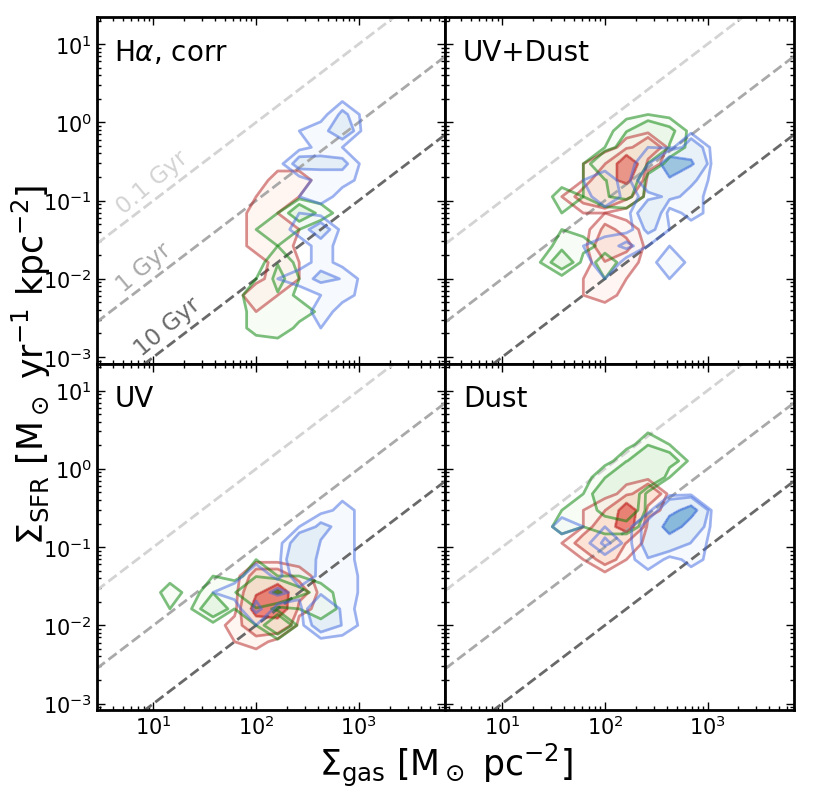}\\
\end{center}
\caption{Star formation rate surface density, computed based on the extinction--corrected H$\alpha$ line, the rest-frame UV luminosity, the dust luminosity, and the sum of UV- and dust-based SFR, in the PJ308--21 system (see Sec.~\ref{sec_ks} for details). The green, blue, and red contours map the quasar host galaxy, the western companion, and the eastern companion respectively. Dashed grey lines mark the loci of constant depletion times.}
\label{fig_KS}
\end{figure}

Fig.~\ref{fig_KS} shows the comparison between the surface distribution of gas and SFR with different tracers in PJ308--21. Diagonal grey lines are loci of constant depletion times $t_{\rm dep}=\Sigma_{\rm gas}/\Sigma_{\rm SFR}$. This is the timescale at which the gas will be depleted, assuming it is consumed at a constant star formation rate. Low values ($t_{\rm dep}\sim 0.1$\,Gyr) are indicative of starburst activity, whereas main sequence galaxies typically show values of $\sim 1$\,Gyr \citep[e.g.,][]{tacconi17,walter20}. The western companion shows higher $\Sigma_{\rm SFR}$(H$\alpha$) and  $\Sigma_{\rm SFR}$(UV) values than the other components, suggesting that it is characterized by a lower fraction of obscured star formation. Conversely, the quasar host galaxy has most of its star formation traced by dust. The quasar host galaxy shows slightly lower values of $t_{\rm dep}$ than the typical main-sequence, but no component of the system appears to reach values typical of starburst conditions ($t_{\rm dep}\sim 0.1$\,Gyr) (with the caveat that we do not probe the central part of the quasar host galaxy here; see footnote 3).

\section{Discussion and conclusions}

We have presented an integral field spectroscopy study at rest-frame optical wavelengths of the $z=6.2342$ quasar PJ308--21 and its close galactic environment. The system is detected and mapped in several emission lines: \Ha{}, \Hb{}, \Oiii{}$_{\rm 5007\,\AA}$, \Oiii{}$_{\rm 4959\,\AA}$, \Nii{}$_{\rm 6584\,\AA}$, \Sii{}$_{\rm 6717\,\AA}$, \Sii{}$_{\rm 6731\,\AA}$, \Heii{}$_{\rm 4686\,\AA}$. This suite of ISM probes enables an unprecedented look into the astrophysics of the ionized gas in this assembling system at the dawn of galaxy formation. Our main findings are the following:
\begin{itemize}
\item[{\em i-}] The system is organized in five components: The quasar host galaxy, the western and eastern companions, the bridge connecting the quasar host with the western companion, and the outflow. All these components are mapped in their \Oiii{} and \Ha{} line emission, while detections of other lines display a range of significance.
\item[{\em ii-}] To first order, the morphology of the system observed in \Ha{} and other lines resembles the \Cii{} map observed with ALMA. Notable differences are: 1) the rest-frame optical line emission are hard to characterize in the central $\lsim 0.3''$ due to the unfavorable contrast with respect to the bright nucleus, whereas \Cii{} observations do not suffer from this limitation; 2) the western companion shows much more prominent line emission at optical wavelengths than what we observed in the far-infrared bands; 3) optical lines observed with JWST appear to miss a more diffuse component that was detected in \Cii{} due to surface brightness limitations; 4) the outflow is detected in the optical \Ha{} and \Oiii{} lines, but not in \Cii{}, suggesting that this is a purely ionized component with no dense, cold clouds.
\item[{\em iii-}] The observed line ratios allow us to map the photoionization conditions in the system. We conclude that the quasar host galaxy experiences photoionization from the quasar. The western companion is photoionized by young stars, while in the eastern companion both in-situ star formation and the nearby quasar may contribute to the photoionization budget. Our observations strongly disfavor an AGN scenario for both the eastern and the western companion galaxies of the quasar host, contrary to previous claims based on X-ray observations \citep{connor19}.
\item[{\em iv-}] The eastern and western companion galaxies show different photoionization, density, and --- critically --- metallicity values. This suggests that these two components are actually two different galaxies, and not a single, tidally stripped galaxy. This points to a high merger rate for quasar host galaxies within the first Gyr of the Universe. The two companion galaxies are rather massive (with stellar masses close to $10^{10}$\,\Msun{}). Their metallicity is lower than solar, but still relatively high compared to typical galaxies at these redshifts \citep[$Z<0.3$\,Z$_\odot$; see, e.g.,][]{taylor22}. This suggests that these galaxies are already well evolved systems. This analysis adds quantitative evidence that the build-up of quasar host galaxies at cosmic dawn happens in an extremely efficient fashion, in the core of prominent galaxy overdensities. 
\end{itemize}

This project showcases the transformational impact of JWST on our understanding of the build-up of the first massive galaxies and on the early growth of massive black holes. Measurements that until recently were beyond reach, can now not only be performed within a few hours of integration, but even mapped on a spatially--resolved scale. As the number of known quasar+companion galaxy systems at $z>6$ is rapidly growing, this work paves the way to future campaigns aimed at characterizing the build-up of early quasar host galaxies in a statistically--significant sample.

\begin{acknowledgements} 
We thank the anonymous referee for their constructive feedback. RD and LF acknowledge support from the INAF GO 2022 grant ``The birth of the giants: JWST sheds light on the build-up of quasars at cosmic dawn''. RD and AL acknowledge support by the PRIN MUR ``2022935STW'', RFF M4.C2.1.1, CUP J53D23001570006 and C53D23000950006. LF acknowledges support from the INAF 2023 mini-grant ``Exploiting the powerful capabilities of JWST/NIRSpec to unveil the distant Universe''. RAM acknowledges support from the Swiss National Science Foundation (SNSF) through project grant 200020\_207349. BT acknowledges support from the European Research Council (ERC) under the European Union’s Horizon 2020 research and  innovation program (grant agreement 950533) and from the Israel Science Foundation (grant 1849/19). MT acknowledges support from the NWO grant 0.16.VIDI.189.162 (``ODIN''). MVe gratefully acknowledges support from the Independent Research Fund Denmark via grant number DFF 8021-00130. 
\end{acknowledgements} 

\begin{table*}
\caption{Line fluxes and velocity shifts for the spectral measurements presented in this work. (1) Fitted line. (2) Rest-frame wavelength of the fitted emission line. (3--7) Measured values for the regions described in Fig.~\ref{fig_system_map}. }\label{tab_spec_fits}
\vspace{-5mm}
\begin{center}
\begin{tabular}{cc|ccccc}
\hline
Line &  $\lambda_0$ [\AA{}] & West Comp. & Bridge & Outflow & Host & East Comp. \\
(1)  & (2)                 & (3)   & (4)   & (5)   & (6)   & (7)    \\
\hline
\multicolumn{7}{c}{$F_{\rm line}$ [$10^{-18}$\,erg\,s$^{-1}$\,cm$^{-2}$]}\\
\hline
H$\alpha$ & $ 6562 $ & $55.04\pm 0.36$     & $7.27\pm 0.33$ & $2.00\pm 0.23$	   & $3.81\pm0.31$	& $14.01\pm 0.35$ \\  
H$\beta$  & $ 4861 $ & $16.50\pm 0.21$     & $2.54\pm 0.16$ & $0.85\pm 0.20$	   & $0.83\pm0.16$	& $ 5.85\pm 0.20$ \\
H$\gamma$ & $ 4340 $ &  $9.16\pm 0.25$     & $1.54\pm 0.19$ & $ <4.15$  	   & $<3.64$		& $ 0.98\pm 0.31$ \\
H$\delta$ & $ 4101 $ &  $4.27\pm 0.30$     & $2.23\pm 0.25$ & $ <0.54$  	   & $<0.68$		& $ 1.56\pm 0.28$ \\
He {\sc i} & $ 4471 $ &  $1.97\pm 0.21$     & $1.16\pm 0.24$ & $ <0.40$  	   & $<0.51$		& $ 0.65\pm 0.21$ \\
\Heii     & $ 4685 $ &  $2.55\pm 0.18$     & $1.27\pm 0.17$ & $ <0.40$  	   & $<0.48$		& $ 2.22\pm 0.22$ \\
\Oiii     & $ 4363 $ &  $1.88\pm 0.21$     & $0.84\pm 0.19$ & $ <0.78$  	   & $<0.61$		& $ 2.62\pm 0.27$ \\
\Oiii     & $ 4959 $ & $37.83\pm 0.19$     & $1.02\pm 0.16$ & $1.184\pm 0.14$	   & $2.21\pm 0.16$	& $ 4.06\pm 0.20$ \\
\Oiii     & $ 5007 $ &$113.40\pm 0.26$     & $6.07\pm 0.25$ & $4.527\pm 0.15$	   & $9.63\pm 0.16$	& $ 9.09\pm 0.22$ \\
He {\sc i} & $ 5875 $ &  $3.63\pm 0.23$     & $1.01\pm 0.23$ & $ <0.48$  	   & $<0.68$		& $ <0.87$	  \\
\Oi{}     & $ 6300 $ &  $ <0.88$	   & $   <0.77$     & $ <0.56$  	   & $<2.44$		& $ <0.98$	  \\
\Nii{}    & $ 6548 $ &  $5.06\pm 0.31$     & $2.83\pm 0.29$ & $ <0.74$  	   & $<0.87$		& $ 8.20\pm 0.35$ \\
\Nii{}    & $ 6583 $ &  $3.17\pm 0.32$     & $1.46\pm 0.34$ & $ <0.70$  	   & $<0.84$		& $ 5.45\pm 0.45$ \\
\Sii{}    & $ 6716 $ &  $2.34\pm 0.34$     & $1.20\pm 0.34$ & $ <0.91$  	   & $<0.93$		& $ 5.57\pm 0.39$ \\
\Sii{}    & $ 6731 $ &  $2.84\pm 0.33$     & $1.84\pm 0.31$ & $ <0.69$  	   & $<0.89$		& $ 2.92\pm 0.44$ \\
\hline
\multicolumn{7}{c}{$\Delta v$ [\kms{}]}\\
\hline
H$\alpha$ & $ 6562 $&	-631   & -704  &  208  &  252	&  530 \\
H$\beta$  & $ 4861 $&	-647   & -917  &  126  &  258	&  443 \\
H$\gamma$ & $ 4340 $&	-647   & -749  & ---   &  ---	& 1268 \\
H$\delta$ & $ 4101 $&	-615   & -981  & ---   &  ---	&  766 \\
He {\sc i} & $ 4471 $&	-623   & -794  &  ---  &  ---	&  681 \\
\Heii     & $ 4685 $&	-765   & -605  &  ---  &  ---	&  562 \\
\Oiii     & $ 4363 $&	-792   & -798  &  ---  &  ---	&  449 \\
\Oiii     & $ 4959 $&	-648   & -978  &   65  &  214	&  561 \\
\Oiii     & $ 5007 $&	-651   & -803  &  113  &  244	&  554 \\
He {\sc i} & $ 5875 $&	-660   & -649  &  ---  &  ---	&  --- \\
\Oi{}     & $ 6300 $&	 ---   &  ---  &  ---  &  ---	&  --- \\
\Nii{}    & $ 6548 $&	-430   &  115  &  ---  &  ---	&  752 \\
\Nii{}    & $ 6583 $&	-502   &-1872  &  ---  &  ---	&  533 \\
\Sii{}    & $ 6716 $&	-624   & -890  &  ---  &  ---	&  621 \\
\Sii{}    & $ 6731 $&	-687   & -574  &  ---  &  ---	&  157 \\
\hline
\end{tabular}
\end{center}
\end{table*}

\begin{appendix}

\section{Detector-based stripes}\label{sec_detector}

\begin{figure}
\begin{center}
\includegraphics[width=0.49\textwidth]{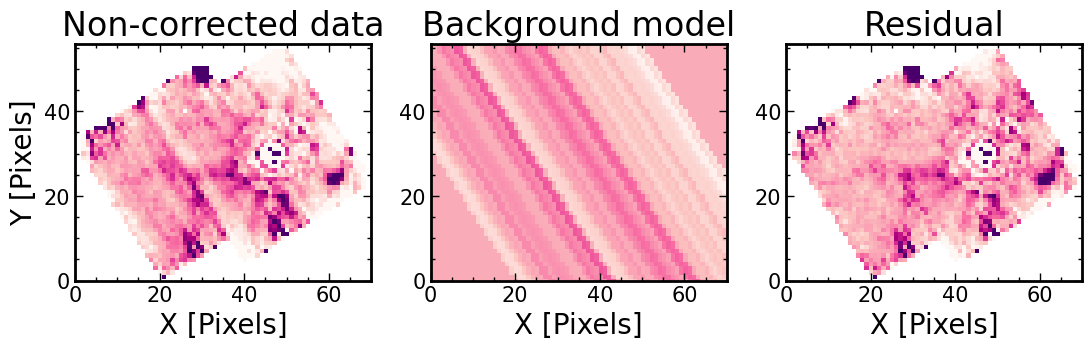}\\
\end{center}
\caption{The detector stripes observed in one of the continuum--only images produced after PSF subtraction, the model created with the procedure described in appendix \ref{sec_detector}, and the residual image used in the analysis.}
\label{fig_detector}
\end{figure}

The pipeline version in use (1.12.5) introduces stripes after drizzling, which are most visible in the continuum images (see Fig.~\ref{fig_detector}). The orientation of the stripes aligns with the position angle of the IFU. The stripes are different in different cubes, and have been reported as well in other datasets (Eros Vanzella et al., private communication). The stripes do not appear to depend on the presence of the signal in the data. We thus opt to treat this as an additive component of the background. We model it by computing the median value of pixels along a 1-pixel wide column tilted as the IFU (based on the position angle information stored in the header keyword \textsf{'PA\_APER'}). This background model is recomputed on a channel-by-channel basis and then subtracted off from the drizzled cubes as well as from the collapsed continuum images. In order to maximize the number of ``pure background'' pixels available for this analysis, we apply this correction after the PSF subtraction described in sec.~\ref{sec_psf}.

\end{appendix}

\end{document}